\documentstyle[epsf]{kapmono}

\kluwerbib

\begin{document}

%
%

\def\BCMO{\rm {Bi_{1-x} Ca_x Mn O_3}}
\def\PSMO{\rm {Pr_{1-x} Sr_x Mn O_3}}
\def\NSMO{\rm {Nd_{1-x} Sr_x Mn O_3}}
\def\PCMO{\rm {Pr_{1-x} Ca_x Mn O_3}}
\def\LCMOhalf{\rm {La_{0.5} Ca_{0.5} Mn O_3}}
\def\LCMO{\rm {La_{1-x} Ca_x Mn O_3}}
\def\LSMO{\rm {La_{1-x} Sr_x Mn O_3}}
\def\bilayered(1.8){\rm {La_{1.2} Sr_{1.8} Mn_2 O_7 }}
\def\bilayer{\rm {La_{2-2x} Sr_{1+2x} Mn_2 O_7 }}
\def\single{\rm {La_{1-x} Sr_{1+x} Mn O_4}}
\def\densi{$\langle$$n$$\rangle$}

\chapter{Theory of Manganites}

\begin{center}
Takashi HOTTA$^1$ and Elbio DAGOTTO$^2$ \\
\end{center}

\begin{center}
$^1$Advanced Science Research Center \\
Japan Atomic Energy Research Institute \\
Tokai, Ibaraki 319-1195, JAPAN
\end{center}

\begin{center}
$^2$National High Magnetic Field Laboratory \\
Florida State University \\
Tallahassee, FL32310, U.S.A.
\end{center}

\medskip

In this review, the present status of theories for manganites
is discussed. The complex phase diagrams of these materials, with
a variety of spin-charge-orbital ordering tendencies, is addressed
using mean-field and Monte Carlo simulation techniques. The stability
of the charge-ordered states, such as the CE-state at half-doping,
appears to originate, in part, in the topology of the zigzag chains
present in those states. In addition, it is argued that phase
separation tendencies are notorious in realistic models for Mn-oxides. 
They produce nanoscale clusters of competing phases, either through 
an electronic separation tendency or through the influence of disorder 
on first-order transitions. These inhomogeneities lead to a ``colossal''
magnetoresistance (CMR) effect, compatible with experiments. This brief 
review is based on a more extensive work recently presented 
[E. Dagotto, T. Hotta, and A. Moreo, Phys. Rep. {\bf 344}, 1 (2001)]. 
There, a comprehensive analysis of the experimental literature can be
found. In real manganites, the tendencies toward inhomogeneous states
are notorious in CMR regimes, in excellent agreement with the
theoretical description outlined here.

\par\vfill
\eject

\section{Early Theoretical Studies of Manganites}

Most of the early theoretical work on manganites focused on the
qualitative aspects of the experimentally discovered relation between
transport and magnetic properties, namely the increase in conductivity
upon the polarization of the spins. 
Not much work was devoted to the magnitude of the magnetoresistance
effect itself.
The formation of coexisting clusters of competing phases was not
included in the early considerations, but this is the dominant theory
at present. 
The states of manganites were assumed to be uniform, and 
``Double Exchange'' (DE) was proposed by Zener (1951) as a way to 
allow for charge to move in manganites by the generation of 
a spin polarized state.
The DE process has been historically explained in two somewhat
different ways.
Originally, Zener (1951) considered the explicit movement of
electrons schematically written as 
$\rm Mn^{3+}_{1\uparrow}$$\rm O_{2\uparrow,3\downarrow}$$\rm Mn^{4+}$
$\rightarrow$
$\rm Mn^{4+}$$\rm O_{1\uparrow,3\downarrow}$$\rm Mn^{3+}_{2\uparrow}$
where 1, 2, and 3 label electrons that belong either to the oxygen
between manganese, or to the $e_{\rm g}$-level of the Mn-ions.
In this process there are two $simultaneous$ motions (thus the name
double-exchange) involving electron 2 moving from the oxygen to the
right Mn-ion, and electron 1 from the left Mn-ion to the oxygen. 
The second way to visualize DE processes was presented in detail by
Anderson and Hasegawa (1955) and it involves a second-order process in
which the two states described above go from one to the other using an
intermediate state $\rm Mn^{3+}_{1\uparrow}$$\rm O_{3\downarrow}$
$\rm Mn^{3+}_{2\uparrow}$.
In this context the effective hopping for the electron to move from
one Mn-site to the next is proportional to the square of the hopping
involving the $p$-oxygen and $d$-manganese orbitals ($t_{\rm pd}$). 
In addition, if the localized spins are considered classical and with
an angle $\theta$ between nearest-neighbor ones, the effective hopping
becomes proportional to $\rm cos(\theta/2)$, as shown by Anderson and
Hasegawa (1955). 
If $\theta$=0 the hopping is the largest, while if $\theta$=$\pi$,
corresponding to an antiferromagnetic background, then the hopping
cancels. 

Note that the oxygen linking the Mn-ions is crucial to understand the
origin of the word ``double'' in this process. 
Nevertheless, the majority of the theoretical work carried out in the
context of manganites simply forgets the
presence of the oxygen and uses a manganese-only Hamiltonian.
It is interesting to observe that ferromagnetic states appear in this
context even $without$ the oxygen.
It is clear that the electrons simply need a polarized background to
improve their kinetic energy, in similar ways as the Nagaoka phase is
generated in the one-band Hubbard model at large $U/t$ 
(for a high-$T_{\rm c}$ review, see Dagotto, 1994). 
This tendency to optimize the kinetic energy is at work in a variety
of models and the term double-exchange appears unnecessary.
However, in spite of this fact it has become customary to refer to
virtually any ferromagnetic phase found in manganese models as 
``DE induced'' or ``DE generated'', forgetting the historical origin
of the term. 
In this review a similar convention will be followed, namely the
credit for the appearance of FM phases will be given to the DE
mechanism, although a more general and simple kinetic-energy
optimization is certainly at work. 

Regarding the stabilization of ferromagnetism, computer simulations
(Yunoki et al., 1998a) and a variety of other approximations have
clearly shown that models $without$ the oxygen degrees of freedom (to
be reviewed below) can also produce FM phases, as long as the Hund
coupling is large enough.
In this situation, when the $e_{\rm g}$ electrons directly jump from
Mn to Mn their kinetic energy is minimized if all spins are aligned. 
As explained before, this procedure to obtain
ferromagnetism is usually also called double-exchange and even the
models from where it emerges are called double-exchange models.

At this point it is useful to discuss the well-known proposed
 ``spin canted'' state for manganites. Work by de Gennes (1960) using 
mean-field approximations suggested that the interpolation between the 
antiferromagnetic state of the undoped limit and the ferromagnetic
state at  finite hole density, where the DE mechanism works, occurs
through a ``canted state'', similar as the state produced by a
magnetic field acting over an antiferromagnet.
In this state the spins develop a moment in one direction, while
being mostly antiparallel within the plane perpendicular to that
moment. The coexistence of FM and AF features in several experiments
carried out at low hole doping (some of them reviewed below) led to
the widely spread belief, until recently, that this spin canted state
was indeed found in real materials.
However, a plethora of recent theoretical work (also discussed below)
has shown that the canted state is actually not realized in the
model of manganites studied by de Gennes (i.e., the simple one-orbital
model). Instead phase separation occurs between the AF and FM states,
as extensively reviewed below. 
Nevertheless, a spin canted state is certainly still a possibility in
real low-doped manganites but its origin, if their presence is
confirmed, needs to be revised.
It may occur that substantial Dzyaloshinskii-Moriya (DM) interactions
appear in manganese oxides, but the authors are not aware of
experimental papers confirming or denying their relevance.

Early theoretical work on manganites carried out by Goodenough (1955)
explained many of the features observed in the neutron scattering
$\LCMO$ experiments by Wollan and Koehler (1955), notably the appearance
of the A-type AF phase at x=0 and the CE-type phase at x=0.5.
The approach of Goodenough (1955) was based on the notion of
``semicovalent exchange''.
Analyzing the various possibilities for the orbital directions and
generalizing to the case where Mn$^{4+}$ ions are also present,
Goodenough (1955) arrived to the A- and CE-type phases of manganites
very early in the theoretical study of these compounds.
In this line of reasoning, note that the Coulomb interactions are
important to generate Hund-like rules and the oxygen is also important 
to produce the covalent bonds. The lattice distortions are also quite 
relevant in deciding which of the many possible states minimizes the
energy. However, it is interesting to observe that in more recent
theoretical work described below in this review, both the A- and
CE-type phases can be generated without the explicit appearance of
oxygens in the models and also without including long-range Coulombic
terms.
 
Summarizing, there appears to be three mechanisms to produce effective
FM interactions: (i) double exchange, where electrons are mobile,
which is valid for non charge-ordered states and where the oxygen 
plays a key role, (ii) Goodenough's approach where covalent bonds are
important (here the electrons do not have mobility in spite of the FM
effective coupling), and it mainly applies to charge-ordered states,
and (iii) the approach based on
purely Mn models (no oxygens) which leads to FM interactions mainly as
a consequence of the large Hund coupling in the system.
If phonons are introduced in the model it can be shown that the A-type
and CE-type states are generated.
In the remaining theoretical part of the review most of the emphasis
will be given to approach (iii) to induce FM bonds since a large
number of experimental results can be reproduced by this procedure,
but it is important to keep in mind the historical path followed in
the understanding of manganites.

Based on all this discussion, it is clear that reasonable proposals to
understand the stabilization of AF and FM phases in manganites have
been around since the early theoretical studies of manganese oxides.
However, these approaches (double exchange, ferromagnetic covalent
bonds, and large Hund coupling) are still $not$ sufficient to handle
the very complex phase diagram of manganites.
For instance, there are compounds such as $\LSMO$ that actually 
do not have the CE-phase at x=0.5, while others do. 
There are compounds that are never metallic, while others have a
paramagnetic state with standard metallic characteristics. 
And even more important, in the early studies of manganites there was
no proper rationalization for the large MR effect. 
It is only with the use of state-of-the-art many-body tools that the
large magnetotransport effects are starting to be understood, thanks to 
theoretical developments in recent years that can address the competition
among the different phases of manganites, their clustering and
mixed-phase tendencies, and dynamical Jahn-Teller polaron
formation.

The prevailing ideas to explain the curious magnetotransport behavior 
of manganites changed in the mid-90's from the simple double-exchange
scenario to a more elaborated picture where a large Jahn-Teller (JT)
effect, which occurs in the Mn$^{3+}$ ions, produces a strong 
electron-phonon coupling that persists even at densities where a
ferromagnetic ground-state is observed.
In fact, in the undoped limit x=0, and even at finite but small x, it
is well-known that a robust static structural distortion is present in
the manganites.
In this context, it is natural to imagine the existence of small
lattice polarons in the paramagnetic phase above $T_{\rm C}$, and it
was believed that these polarons lead to the insulating behavior of
this regime. Actually, the term polaron 
is somewhat ambiguous.
In the context of manganites it is usually associated with a local
distortion of the lattice around the charge, sometimes
together with a magnetic cloud or region with ferromagnetic 
correlations (magneto polaron or lattice-magneto polaron).

The fact that double-exchange cannot be enough to understand the
physics of manganites is clear from several different points of view.
For instance, Millis, Littlewood and Shraiman (1995) arrived at this
conclusion by presenting estimations of the critical Curie temperature
and of the resistivity using the DE framework.
It is clear that the one-orbital model is incomplete for
quantitative studies since it cannot describe, e.g., the key
orbital-ordering of manganites and the proper charge-order states at x
near 0.5, which are so important for the truly CMR effect found in
low-bandwidth manganites. 
Not even a fully disordered set of classical spins
can scatter electrons as much as needed to reproduce the experiments
(again, unless large antiferromagnetic regions appear in a mixed-phase
regime).

Millis, Shraiman and Mueller (1996) (see also
Millis, Mueller, and  Shraiman, 1996, and Millis, 1998)
argued that the physics of
manganites is dominated by the interplay between a strong
electron-phonon coupling and the large Hund coupling effect that
optimizes the electronic kinetic energy by the generation of a FM
phase. 
The large value of the electron phonon coupling is clear in the regime 
of manganites below x=0.20 where a static JT distortion plays a key
role in the physics of the material. Millis, Shraiman and Mueller
(1996) argued that a dynamical JT effect may persist at higher hole
densities, without leading to long-range order but producing important
fluctuations that localize electrons by splitting the degenerate $e_{\rm g}$
levels at a given MnO$_6$ octahedron. 
The calculations were carried out using the infinite dimensional
approximation that corresponds to a $local$ mean-field technique where
the polarons can have only a one site extension, and the classical
limit for the phonons and spins was used. 
The Coulomb interactions were neglected, but further work reviewed
below showed that JT and Coulombic interactions lead to very similar
results (Hotta, Malvezzi, and Dagotto, 2000).
Millis, Shraiman and Mueller (1996) argued that the ratio 
$\lambda_{\rm eff}$=$E_{\rm JT}$/$t_{\rm eff}$ dominates the physics
of the problem. Here $E_{\rm JT}$ is the static trapping energy 
at a given octahedron, and $t_{\rm eff}$ is an effective hopping that
is temperature dependent following the standard DE discussion. 
In this context it was conjectured that when the temperature is larger
than $T_{\rm C}$ the effective coupling $\lambda_{\rm eff}$ could be
above the critical value that leads to insulating behavior due to
electron localization, while it becomes smaller than the critical
value below $T_{\rm C}$, thus inducing metallic behavior.
However, in order to describe the percolative nature of the transition 
found experimentally and the notorious phase separation tendencies,
calculations beyond mean-field approximations are needed, as reviewed
later in this paper. Similar phase separation ideas have been discussed
extensively in the context of high temperature superconductors as well
(see Emery, Kivelson, and Lin, 1990. See also Tranquada et al., 1995).

\section{Model for Manganites}

Before proceeding to a description of the latest theoretical developments,
it is necessary to clearly write down the model Hamiltonian for manganites.
For complex materials such as the Mn-oxides, unfortunately, the full
Hamiltonian includes several competing tendencies
and couplings.
However, as shown below, the essential physics can be obtained using
relatively simple models, deduced from the complicated full
Hamiltonian.

\subsection{Crystal field effect}

In order to construct the model Hamiltonian for manganites, let us
start our discussion at the level of the atomic problem, in which just 
one electron occupies a certain orbital in the $3d$ shell of a
manganese ion.
Although for an isolated ion a five-fold degeneracy exists for the
occupation of the $3d$ orbitals, this degeneracy is partially lifted
by the crystal field due to the six oxygen ions surrounding the
manganese forming an octahedron.
This is analyzed by the ligand field theory that shows that the
five-fold degeneracy is lifted into doubly-degenerate 
$e_{\rm g}$-orbitals ($d_{x^2-y^2}$ and $d_{3z^2-r^2}$) and
triply-degenerate $t_{\rm 2g}$-orbitals ($d_{xy}$, $d_{yz}$, and
$d_{zx}$).
The energy difference between those two levels is usually expressed as 
$10Dq$, by following the traditional notation in the ligand field
theory.

Here note that the energy level for the $t_{\rm 2g}$-orbitals is lower 
than that for $e_{\rm g}$-orbitals.
Qualitatively this can be understood as follows:
The energy difference originates in the Coulomb interaction
between the $3d$ electrons and the oxygen ions surrounding manganese.
While the wave-functions of the $e_{\rm g}$-orbitals is extended along
the direction of the bond between manganese and oxygen ions, those in
the $t_{\rm 2g}$-orbitals avoid this direction.
Thus, an electron in $t_{\rm 2g}$-orbitals is not heavily influenced
by the Coulomb repulsion due to the negatively charged oxygen ions, 
and the energy level for $t_{\rm 2g}$-orbitals is lower than that
for $e_{\rm g}$-orbitals. 

As for the value of $10Dq$, it is explicitly written as
(see Gerloch and Slade, 1973)
\begin{equation}
  10Dq=\frac{5}{3} \frac {Ze^2}{a}
  \frac{\langle r^4 \rangle}{a^4},
\end{equation}
where $Z$ is the atomic number of the ligand ion, $e$ is the electron 
charge, $a$ is the distance between manganese and oxygen ions, 
$r$ is the coordinate of the $3d$ orbital, and
${\langle \cdots \rangle}$ denotes the average value by using the
radial wavefunction of the $3d$ orbital.
Estimations by Yoshida (page 29 of Yoshida, 1996) suggest that 10$Dq$ is
about 10000-15000cm$^{-1}$ (remember that 1eV = 8063 cm$^{-1}$).

\subsection{Coulomb interactions}

Now consider a Mn$^{4+}$ ion, in which three electrons exist
in the $3d$ shells.
Although those electrons will occupy $t_{\rm 2g}$-orbitals due to the
crystalline field splitting, the configuration is not uniquely
determined.
To configure three electrons appropriately, it is necessary to take 
into account the effect of the Coulomb interactions.
In the localized ion system, the Coulomb interaction term
among $d$-electrons is generally given by
\begin{eqnarray}
  \label{eq:Coulomb0}
  H^{\rm C}_{\bf i} \!=\! \frac{1}{2}
  \sum_{\gamma_1 \gamma_2 \gamma'_1 \gamma'_2}
  \sum_{\sigma_1 \sigma_2 \sigma'_1 \sigma'_2}
  I_{\gamma_1 \sigma_1,\gamma_2 \sigma_2;
  \gamma'_1 \sigma'_1,\gamma'_2 \sigma'_2}
  d_{{\bf i}\gamma_1\sigma_1}^{\dag}  
  d_{{\bf i}\gamma_2\sigma_2}^{\dag}
  d_{{\bf i}\gamma'_2\sigma'_2}
  d_{{\bf i}\gamma'_1\sigma'_1},
\end{eqnarray}
where $d_{{\bf i}\gamma\sigma}$ is the annihilation operator for a 
$d$-electron with spin $\sigma$ in the 
$\gamma$-orbital at site ${\bf i}$,
and the Coulomb matrix element is given by
\begin{eqnarray}
  \label{element}
  \! I_{\gamma_1 \sigma_1,\gamma_2 \sigma_2;
  \gamma'_1 \sigma'_1,\gamma'_2 \sigma'_2}
  \!=\! \int \! \int \! d{\bf r} d{\bf r'} \!
  \phi_{\gamma_1\sigma_1}^{*}\!({\bf r})
  \phi_{\gamma_2\sigma_2}^{*}\!({\bf r'})
  g_{{\bf r}-{\bf r'}}
  \phi_{\gamma'_1\sigma'_1}\!({\bf r})
  \phi_{\gamma'_2\sigma'_2}\!({\bf r'}).
\end{eqnarray}
Here $g_{{\bf r}-{\bf r'}}$ is the screened Coulomb potential,
and $\phi_{\gamma\sigma}({\bf r})$ is the Wannier function for an electron
with spin $\sigma$ in the $\gamma$-orbital at position ${\bf r}$.
By using the Coulomb matrix element, the so-called ``Kanamori parameters'', 
$U$, $U'$, $J$, and $J'$, are defined as follows
(see Kanamori, 1963; Dworin and Narath, 1970;
Castellani et al., 1978).
$U$ is the intraband Coulomb interaction, given by 
\begin{eqnarray}
  U = I_{\gamma \sigma,\gamma \sigma';
      \gamma \sigma,\gamma \sigma'},
\end{eqnarray}
with $\sigma \ne \sigma'$.
$U'$ is the interband Coulomb interaction, expressed by 
\begin{eqnarray}
  U'= I_{\gamma \sigma,\gamma' \sigma';
  \gamma \sigma,\gamma' \sigma'},
\end{eqnarray}
with $\gamma \ne \gamma'$.
$J$ is the interband exchange interaction, written as 
\begin{eqnarray}
  J = I_{\gamma \sigma,\gamma' \sigma';
  \gamma' \sigma,\gamma \sigma'},
\end{eqnarray}
with $\gamma \ne \gamma'$.
Finally, $J'$ is the pair-hopping amplitude
between different orbitals, given by
\begin{eqnarray}
  J' = I_{\gamma \sigma,\gamma \sigma';
  \gamma' \sigma,\gamma' \sigma'},
\end{eqnarray}
with $\gamma \ne \gamma'$ and $\sigma \ne \sigma'$.

Note the relation $J$=$J'$, which is simply due to the fact that each of 
the parameters above is given by an integral of the Coulomb interaction
sandwiched with appropriate orbital wave functions. 
Analyzing the form of those integrals the equality between $J$ and $J'$
can be deduced [see Eq.~(2.6) of Castellani et al., 1978;
See also the Appendix of Fr\'esard and Kotliar, 1997].
Using the above parameters, it is convenient to rewrite
the Coulomb interaction term in the following form:
\begin{eqnarray}
  \label{eq:coulomb}
  H^{\rm C}_{\bf i} &=&
  \frac{U}{2} \sum_{\gamma,\sigma \ne \sigma'}
   n_{{\bf i}\gamma\sigma}n_{{\bf i}\gamma\sigma'}
  + \frac{U'}{2}  \sum_{\sigma,\sigma',\gamma \ne \gamma'} 
  n_{{\bf i}\gamma\sigma}n_{{\bf i}\gamma'\sigma'}
  \nonumber \\
  &+& \frac{J}{2} \sum_{\sigma,\sigma',\gamma \ne \gamma'} 
  d_{{\bf i}\gamma\sigma}^{\dag}  
  d_{{\bf i}\gamma'\sigma'}^{\dag}
  d_{{\bf i}\gamma\sigma'}
  d_{{\bf i}\gamma'\sigma} \nonumber \\
  &+& \frac{J'}{2} \sum_{\sigma \ne \sigma',\gamma \ne \gamma'} 
  d_{{\bf i}\gamma\sigma}^{\dag}  
  d_{{\bf i}\gamma\sigma'}^{\dag}
  d_{{\bf i}\gamma'\sigma'}
  d_{{\bf i}\gamma'\sigma},
\end{eqnarray}
where
$n_{{\bf i}\gamma\sigma}$=
$d_{{\bf i}\gamma\sigma}^{\dag}d_{{\bf i}\gamma\sigma}$.
Here it is important to clarify that the parameters $U$, $U'$, and $J$
are not independent (here $J$=$J'$ is used).
The relation among them in the localized ion problem
has been clarified by group theory arguments,
showing that all the above Coulomb interactions
can be expressed by the so-called ``Racah parameters''
(for more details, see Griffith, 1961.
See also Tang, Plihal, and Mills, 1998).
By using those expressions for $U$, $U'$ and $J$,
it is easily checked that the relation
\begin{equation}
  \label{relation}
  U=U'+2J
\end{equation}
holds in any combination of orbitals.
Note that this relation is needed to recover the rotational invariance
in orbital space.
For more details the reader should consult Dagotto, Hotta, and Moreo, 2001.

Now let us move to the discussion of the configuration of three electrons
for the Mn$^{4+}$ ion.
Since the largest energy scale among the several Coulombic interactions
is $U$, the orbitals are not doubly occupied by both up- and down-spin
electrons.
Thus, only one electron can exist in each orbital of the triply degenerate
$t_{\rm 2g}$ sector.
Furthermore, in order to take advantage of $J$, the spins of those 
three electrons point along the same direction.
This is the so-called ``Hund's rule''.

By adding one more electron to Mn$^{4+}$ with three up-spin
$t_{\rm 2g}$-electrons, let us consider the configuration for the
Mn$^{3+}$ ion.
Note here that there are two possibilities due to the balance between
the crystalline-field splitting and the Hund coupling:
One is the ``high-spin state'' in which an electron occupies the
$e_{\rm g}$-orbital with up spin if the Hund coupling is dominant.
In this case, the energy level appears at $U'$$-$$J$+10$Dq$.
Another is the ``low-spin state'' in which one of the
$t_{\rm 2g}$-orbitals is occupied with a down-spin electron, when the 
crystalline-field splitting is much larger than the Hund coupling.
In this case, the energy level occurs at $U$+$2J$.
Thus, the high spin state appears if $10Dq$$<$$5J$ holds.
Since $J$ is a few eV and $10Dq$ is about 1eV in the manganese oxide,
the inequality $10Dq$$<$$5J$ is considered to hold.
Namely, in the Mn$^{3+}$ ion, the high spin state is realized.

In order to simplify the model without loss of essential physics,
it is reasonable to treat the three spin-polarized
$t_{\rm 2g}$-electrons as a localized ``core-spin'' expressed by 
${\bf S}_{\bf i}$ at site ${\bf i}$, since the overlap integral
between $t_{\rm 2g}$ and oxygen $p\sigma$ orbital is small compared to
that between $e_{\rm g}$ and $p\sigma$ orbitals.
Moreover, due to the large value of the total spin $S$=$3/2$, it is
usually approximated by a classical spin (this approximation will be tested
later using computational techniques).
Thus, the effect of the strong Hund coupling between the 
$e_{\rm g}$-electron spin and localized $t_{\rm 2g}$-spins is
considered by introducing 
\begin{equation}
  H_{\rm Hund} = -J_{\rm H} \sum_{\bf i}
  {\bf s}_{\bf i} \cdot {\bf S}_{\bf j},
\end{equation}
where ${\bf s}_{\bf i}$=
$\sum_{\gamma\alpha\beta}d^{\dag}_{{\bf i}\gamma\alpha}
\sigma_{\alpha\beta}d_{{\bf i}\gamma\beta}$,
$J_{\rm H}$($>$0) is the Hund coupling between
localized $t_{\rm 2g}$-spin and mobile $e_{\rm g}$-electron, 
and ${\sigma}$=$(\sigma_x, \sigma_y, \sigma_z)$ are the Pauli
matrices. The magnitude of $J_{\rm H}$ is of the order of $J$.
Here note that ${\bf S}_{\bf i}$ is normalized as
$|{\bf S}_{\bf i}|$=1. 
Thus, the direction of the classical $t_{\rm 2g}$-spin at site 
${\bf i}$ is defined as 
\begin{equation}
  {\bf S}_{\bf i}=(\sin\theta_{\bf i}\cos\phi_{\bf i}, 
  \sin\theta_{\bf i}\sin\phi_{\bf i}, 
  \cos\theta_{\bf i}),
\end{equation}
by using the polar angle $\theta_{\bf i}$ and the azimuthal angle 
$\phi_{\bf i}$.

Unfortunately, the effect of the Coulomb interaction is not fully
taken into account only by $H_{\rm Hund}$ since there remains the 
direct electrostatic repulsion between $e_{\rm g}$-electrons, 
which will be referred to as the ``Coulomb interaction'' hereafter.
Then, the following term should be added to the Hamiltonian.
\begin{equation}
  H_{\rm el-el} = \sum_{{\bf i}} H_{\bf i}^{\rm C}
  + V \sum_{\langle {\bf i,j} \rangle}
  \rho_{\bf i} \rho_{\bf j},
\end{equation}
where $\rho_{\bf i}=\sum_{\gamma\sigma}n_{{\bf i}\gamma\sigma}$.
Note here that in this expression, the index $\gamma$ for the orbital
degree of freedom runs only in the $e_{\rm g}$-sector.
Note also that in order to consider the effect of the long-range 
Coulomb repulsion between $e_{\rm g}$-electrons, the term including
$V$ is added, where $V$ is the nearest-neighbor Coulomb interaction. 

\subsection{Electron-phonon coupling}

Another important ingredient in manganites is the lattice distortion 
coupled to the $e_{\rm g}$-electrons.
In particular, the double degeneracy in the $e_{\rm g}$-orbitals is
lifted by the Jahn-Teller distortion of the MnO$_6$ octahedron
(Jahn and Teller, 1937).
The basic formalism for the study of electrons coupled to Jahn-Teller
modes has been set up by Kanamori, 1960.
He focused on cases where the electronic orbitals are degenerate in
the undistorted crystal structure, as in the case of Mn in an 
octahedron of oxygens.
As explained by Kanamori, the Jahn-Teller effect in this context
can be simply stated as follows;
when a given electronic level of a
cluster is degenerate in a structure of high symmetry, this structure
is generally unstable, and the cluster will present a
distortion toward a lower symmetry ionic arrangement.
In the case of Mn$^{3+}$, which is doubly degenerate when the crystal 
is undistorted, a splitting will occur when the crystal is distorted. 
The distortion of the MnO$_6$ octahedron is ``cooperative'' since once 
it occurs in a particular octahedron, it will affect the neighbors. 
The basic Hamiltonian to describe the interaction between electrons
and Jahn-Teller modes was written by Kanamori and it is of the form
\begin{equation}
  H^{\rm JT}_{\bf i} = g (Q_{2{\bf i}} T^x_{\bf i} +
  Q_{3{\bf i}} T^z_{\bf i})
  + (k_{\rm JT}/2)(Q_{2{\bf i}}^2+Q_{3{\bf i}}^2),
\end{equation}
where 
$g$ is the coupling constant between the $e_{\rm g}$-electrons and
distortions of the MnO$_6$ octahedron,
$Q_{2{\bf i}}$ and $Q_{3{\bf i}}$ are normal modes of vibration of the 
oxygen octahedron that remove the degeneracy between the electronic
levels, and $k_{\rm JT}$ is the spring constant for the Jahn-Teller
mode distortions.
The pseudospin operators are defined as
\begin{equation}
  T^x_{{\bf i}} \!=\!
  \sum_{\sigma}
  (d_{{\bf i}{\rm a}\sigma}^{\dag}d_{{\bf i}{\rm b}\sigma}
  +d_{{\bf i}{\rm b}\sigma}^{\dag}d_{{\bf i}{\rm a}\sigma}),~~
  T^z_{{\bf i}} \!=\!
  \sum_{\sigma}
  (d_{{\bf i}{\rm a}\sigma}^{\dag}d_{{\bf i}{\rm a}\sigma}
  -d_{{\bf i}{\rm b}\sigma}^{\dag}d_{{\bf i}{\rm b}\sigma}).
\end{equation}
In the expression of $H^{\rm JT}_{\bf i}$, a $T^y_{\bf i}$-term does
not appear for symmetry reasons, since it belongs to 
the $A_{\rm 2u}$ representation.
The non-zero terms should correspond to the irreducible
representations included in $E_{\rm g}$$\times$$E_{\rm g}$, namely,
$E_{\rm g}$ and $A_{\rm 1g}$. 
The former representation is expressed by using the pseudo spin
operators $T^x_{\bf i}$ and $T^z_{\bf i}$ as discussed here, while the
latter, corresponding to the breathing mode, is discussed later.

\begin{figure}[t]
\centerline{\epsfxsize=4.truecm \epsfbox{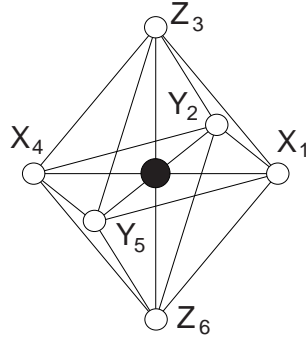}}
\caption{MnO$_6$ octahedron at site ${\bf i}$. 
The labeling for oxygen ions is shown.}
\label{fig1}
\end{figure}

Following Kanamori, $Q_{2{\bf i}}$ and $Q_{3{\bf i}}$ are explicitly
given by
\begin{equation}
  \label{eq:q2}
  Q_{2{\bf i}}={1 \over \sqrt{2}}(X_{1{\bf i}}-X_{4{\bf i}}
  -Y_{2{\bf i}}+Y_{5{\bf i}}),
\end{equation}
and 
\begin{equation}
  \label{eq:q3}
  Q_{3{\bf i}}\!=\! {1 \over \sqrt{6}}(2Z_{3{\bf i}}-2Z_{6{\bf i}}-
  X_{1{\bf i}}+X_{4{\bf i}}-Y_{2{\bf i}}+Y_{5{\bf i}}),
\end{equation}
where $X_{\mu j}$, $Y_{\mu j}$, and $Z_{\mu j}$ are the displacement
of oxygen ions from the equilibrium positions along the $x$-, $y$-,
and $z$-direction, respectively. The convention for the labeling $\mu$
of coordinates is shown in Fig.~\ref{fig1}.

To solve this Hamiltonian,
it is convenient to scale the phononic degrees of freedom as
\begin{equation}
  Q_{2{\bf i}}=(g/k_{\rm JT})q_{2{\bf i}},~~
  Q_{3{\bf i}}=(g/k_{\rm JT})q_{3{\bf i}},
\end{equation}
where $g/k_{\rm JT}$ is the typical length scale for the Jahn-Teller 
distortion, which is of the order of 0.1$\rm \AA$, namely, 2.5\% of
the lattice constant.
When the Jahn-Teller distortion is expressed in the polar coordinate as 
\begin{equation}
  \label{eq:polar}
  q_{2{\bf i}} = q_{{\bf i}} \sin \xi_{\bf i},~~
  q_{3{\bf i}} = q_{{\bf i}} \cos \xi_{\bf i},
\end{equation}
the ground state is easily obtained as
$(-\sin [\xi_{\bf i}/2] {d}_{{\bf i} a\sigma}^{\dag}
+\cos [\xi_{\bf i}/2] {d}_{{\bf i} b\sigma}^{\dag})|0\rangle$
with the use of the phase $\xi_{\bf i}$. 
The corresponding eigenenergy is given by $-E_{\rm JT}$,
where $E_{\rm JT}$ is the static Jahn-Teller energy,
defined by 
\begin{equation}
  E_{\rm JT}=g^2/(2k_{\rm JT}).
\end{equation}
Note here that the ground state energy is independent of the phase 
$\xi_{\bf i}$. Namely, the shape of the deformed isolated octahedron
is not uniquely determined in this discussion.
In the Jahn-Teller crystal, the kinetic motion of $e_{\rm g}$
electrons, as well as the cooperative effect between adjacent
distortions, play a crucial role in lifting the degeneracy and fixing
the shape of the local distortion.

To complete the electron-phonon coupling term, it is necessary to 
consider the breathing mode distortion, coupled to the local electron 
density as
\begin{equation}
  H^{\rm br}_{\bf i} = g Q_{1{\bf i}} \rho_{{\bf i}}
  + (1/2) k_{\rm br}Q_{1{\bf i}}^2,
\end{equation}
where the breathing-mode distortion $Q_{1{\bf i}}$ is given by
\begin{equation}
  \label{eq:q1}
  Q_{1{\bf i}}={1 \over \sqrt{3}}(X_{1{\bf i}}-X_{4{\bf i}}
  +Y_{2{\bf i}}-Y_{5{\bf i}}+Z_{3{\bf i}}-Z_{6{\bf i}}),
\end{equation}
and $k_{\rm br}$ is the associated spring constant.
Note that, in principle, the coupling constants of the $e_{\rm g}$ 
electrons with the $Q_1$, $Q_2$, and $Q_3$ modes could be 
different from one another. For simplicity, here it is assumed that
those coupling constants take the same value.
On the other hand, for the spring constants, a different notation for
the breathing mode is introduced, since the frequency for the
breathing mode distortion has been found experimentally to be
different from that for the Jahn-Teller mode.
This point will be briefly discussed later. 
Note also that the Jahn-Teller and breathing modes are competing with
each other. As it was  shown above, the energy gain due to the
Jahn-Teller distortion is maximized when one electron exists per
site. On the other hand, the breathing mode distortion energy is
proportional to the total number of $e_{\rm g}$ electrons per site,
since this distortion gives rise to an effective on-site attraction
between electrons. 

By combining the Jahn-Teller mode and breathing mode distortions, the
electron-phonon term is summarized as
\begin{equation}
 H_{\rm el-ph}= \sum_{\bf i}(H_{\bf i}^{\rm JT}+H_{\bf i}^{\rm br}).
\end{equation}
This expression depends on the parameter $\beta$=$k_{\rm br}/k_{\rm JT}$,
which regulates which distortion, the Jahn-Teller or breathing 
mode, play a more important role. This point will be discussed 
in a separate subsection.

Note again that the distortions at each site are not independent,
since all oxygens are shared by neighboring MnO$_6$ octahedra,
as easily understood by the explicit expressions of $Q_{1{\bf i}}$,
$Q_{2{\bf i}}$, and $Q_{3{\bf i}}$ presented before.
A direct and simple way to consider this cooperative effect is
to determine the oxygen positions $X_{1{\bf i}}$, $X_{4{\bf i}}$,
$Y_{2{\bf i}}$, $Y_{5{\bf i}}$, $Z_{3{\bf i}}$, and $Z_{6{\bf i}}$,
by using, for instance, the Monte Carlo simulations or numerical
relaxation methods (see Press et al., 1992, chapter 10).
To reduce the burden on the numerical calculations,
the displacements of oxygen ions are assumed to be along the bond
direction between nearest neighboring manganese ions.
In other words, the displacement of the oxygen ion perpendicular to 
the Mn-Mn bond, i.e., the buckling mode, is usually ignored. 
As shown later, even in this simplified treatment, several
interesting results haven been obtained for the spin, charge, and
orbital ordering in manganites.

Rewriting Eqs.~(\ref{eq:q2}), (\ref{eq:q3}), and (\ref{eq:q1})
in terms of the displacement of oxygens from the equilibrium
positions, it can be shown that
\begin{equation}
  Q_{1{\bf i}}=Q_{1}^{(0)}+
  {1 \over \sqrt{3}}(\Delta_{\bf xi}+\Delta_{\bf yi}+\Delta_{\bf zi}), 
\end{equation}
\begin{equation}
  Q_{2{\bf i}}=Q_{2}^{(0)}+
  {1 \over \sqrt{2}}(\Delta_{\bf xi}-\Delta_{\bf yi}),
\end{equation}
\begin{equation}
  Q_{3{\bf i}}=Q_{3}^{(0)}+
  {1 \over \sqrt{6}}(2\Delta_{\bf zi}-\Delta_{\bf xi}-\Delta_{\bf yi}),
\end{equation}
where $\Delta_{\bf ai}$ is given by
\begin{equation}
  \Delta_{\bf ai}=u_{\bf i}^{\bf a}-u_{\bf i-a}^{\bf a},
\end{equation}
with $u_{\bf i}^{\bf a}$ being the displacement of oxygen ion at site
${\bf i}$ from the equilibrium position along the ${\bf a}$-axis.
The offset values for the distortions, $Q_{1}^{(0)}$, $Q_2^{(0)}$, and
$Q_3^{(0)}$, are respectively given by
\begin{equation}
  Q_{1}^{(0)}\!=\!
  {1 \over \sqrt{3}}(\delta L_{\bf x}+\delta L_{\bf y}+\delta L_{\bf z}),
\end{equation}
\begin{equation}
  Q_{2}^{(0)}\!=\!
  {1 \over \sqrt{2}}(\delta L_{\bf x}-\delta L_{\bf y}),
\end{equation}
\begin{equation}  
  Q_{3}^{(0)}\!=\!
  {1 \over \sqrt{6}}(2\delta L_{\bf z}-\delta L_{\bf x}-\delta L_{\bf y}),
\end{equation}
where $\delta L_{\bf a}$=$L_{\bf a}-L$,
the non-distorted lattice constants are $L_{\bf a}$,
 and $L$=$(L_{\bf x}+L_{\bf y}+L_{\bf z})/3$.
In the $cooperative$ treatment,
the $\{u\}$'s  are directly optimized in the numerical calculations
(see Allen and Perebeinos, 1999; Hotta et al. 1999).
On the other hand, in the $non$-$cooperative$ calculations,
$\{Q\}$'s are treated instead of the $\{u\}$'s.
In the simulations, variables are taken as
$\{Q\}$'s or $\{u\}$'s,
depending on the treatments of lattice distortion.

\subsection{Hopping amplitudes}

Although it is assumed that the $t_{\rm 2g}$-electrons are localized
to form core spins, 
the $e_{\rm g}$-electrons can move around the system via the oxygen
$2p$ orbital. 
This hopping motion of $e_{\rm g}$-electrons is expressed as 
\begin{equation}
  H_{\rm kin} =-\sum_{{\bf ia}\gamma \gamma'\sigma}
  t^{\bf a}_{\gamma \gamma'} d_{{\bf i} \gamma \sigma}^{\dag}
  d_{{\bf i+a} \gamma' \sigma},
\end{equation}
where ${\bf a}$ is the vector connecting nearest-neighbor sites and 
$t^{\bf a}_{\gamma \gamma'}$ is the nearest-neighbor hopping amplitude
between $\gamma$- and $\gamma'$-orbitals along the 
${\bf a}$-direction.

The amplitudes are evaluated from the overlap integral between
manganese and oxygen ions by following Slater and Koster (1954).
The overlap integral between $d_{x^2-y^2}$- and $p_x$-orbitals 
is given by
\begin{equation}
  E_{{\bf x},{\rm a}}(\ell,m,n)=(\sqrt{3}/2) \ell
  (\ell^2-m^2)(pd\sigma), 
\end{equation}
where $(pd\sigma)$ is the overlap integral between the $d\sigma$- and
$p\sigma$-orbital and $(\ell,m,n)$ is the unit vector along the
direction from manganese to oxygen ions.
The overlap integral between $d_{3z^2-r^2}$- and $p_x$-orbitals is
expressed as
\begin{equation}
  E_{{\bf x},{\rm b}}(\ell,m,n)= \ell [n^2-(\ell^2+m^2)/2](pd\sigma).
\end{equation}
Thus, the hopping amplitude between adjacent manganese ions along the
$x$-axis via the oxygen $2p_x$-orbitals is evaluated as
\begin{equation}
  -t_{\gamma\gamma'}^{\bf x}=E_{{\bf x},\gamma}(1,0,0) \times 
  E_{{\bf x},\gamma'}(-1,0,0).
\end{equation}
Note here that the minus sign is due to the definition of hopping amplitude
in $H_{\rm kin}$. 
Then, $t_{\gamma\gamma'}^{\bf x}$ is explicitly given by
\begin{equation}
  t_{\rm aa}^{\bf x}
  =-\sqrt{3}t_{\rm ab}^{\bf x}
  =-\sqrt{3}t_{\rm ba}^{\bf x}
  =3t_{\rm bb}^{\bf x}=3t_0/4,
\end{equation}
where $t_0$ is defined by $t_0=(pd\sigma)^2/|\varepsilon_{\rm d}-\varepsilon_{\rm p}|$.
Here $\varepsilon_{\rm d}$ and $\varepsilon_{\rm p}$ are the energy level for
$d$- and $p$-orbitals.
By using the same procedure, the hopping amplitude along the $y$- 
and $z$-axis are given by
\begin{equation}
  t_{\rm aa}^{\bf y}
  =\sqrt{3}t_{\rm ab}^{\bf y}
  =\sqrt{3}t_{\rm ba}^{\bf y}
  =3t_{\rm bb}^{\bf y}=3t_0/4,
\end{equation}
and
\begin{equation}
  t_{\rm bb}^{\bf z}=t_0,
  t_{\rm aa}^{\bf z}=t_{\rm ab}^{\bf z}=t_{\rm ba}^{\bf z}=0,
\end{equation}
respectively.
It should be noted that the signs in the hopping amplitudes between
different orbitals are different between the $x$- and $y$-directions,
which will be important when the charge-orbital ordered phase
in the doped manganites is considered.
Note also that in some cases, it is convenient to define 
$t_{\rm aa}^{\bf x}$ as the energy scale $t$, given as $t$=$3t_0/4$.

\subsection{Heisenberg term}

Thus far, the role of the $e_{\rm g}$-electrons has been discussed
to characterize the manganites.
However, in the fully hole-doped manganites composed of Mn$^{4+}$ 
ions, for instance CaMnO$_3$, it is well known that a G-type 
antiferromagnetic phase appears,
and this property cannot be understood within the above discussion. 
The minimal term to reproduce this antiferromagnetic property
is the Heisenberg-like coupling between localized $t_{\rm 2g}$ spins,
given in the form 
\begin{equation}
  H_{\rm AFM} = J_{\rm AF} \sum_{\langle {\bf i,j} \rangle}
  {\bf S}_{\bf i} \cdot {\bf S}_{\bf j},
\end{equation}
where $J_{\rm AF}$ is the AFM coupling between nearest neighbor 
$t_{\rm 2g}$ spins.
The existence of this term is quite natural from the viewpoint of the
super-exchange interaction, working between neighboring localized
$t_{\rm 2g}$-electrons.
As for the magnitude of $J_{\rm AF}$, it is discussed later
in the text. 

\subsection{Several Hamiltonians}

\begin{figure}[t]
\centerline{\epsfxsize=11.truecm \epsfbox{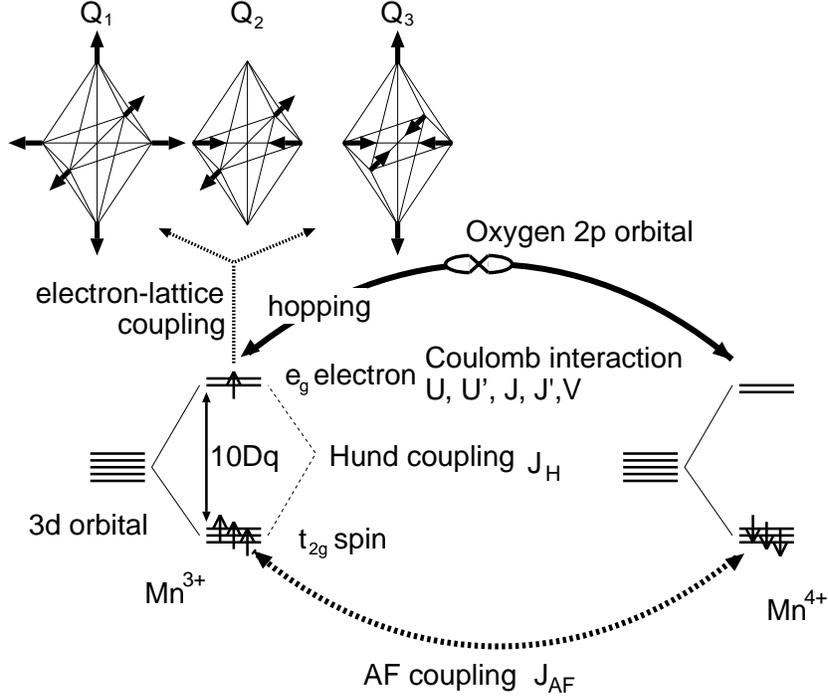}}
\caption{Schematic figure of the model Hamiltonian for manganites.}
\label{fig2}
\end{figure}

As discussed in the previous subsections, there are five important
ingredients that regulate the physics of electrons in manganites:
(i) $H_{\rm kin}$, the kinetic term of the $e_{\rm g}$ electrons. 
(ii) $H_{\rm Hund}$, the Hund coupling between the $e_{\rm g}$
electron spin and the localized $t_{\rm 2g}$ spin.
(iii) $H_{\rm AFM}$, the AFM Heisenberg coupling between nearest
neighbor $t_{\rm 2g}$ spins.
(iv) $H_{\rm el-ph}$, the coupling between the $e_{\rm g}$ electrons
and the local distortions of the MnO$_6$ octahedron.
(v) $H_{\rm el-el}$, the Coulomb interactions among the $e_{\rm g}$
electrons.
As schematically summarized in Fig.~\ref{fig2},
by unifying those five terms
into one, the full Hamiltonian $H$ is defined as
\begin{equation}
  \label{Hamiltonian}
  H = H_{\rm kin} + H_{\rm Hund} + H_{\rm AFM} 
  + H_{\rm el-ph} + H_{\rm el-el}.
\end{equation}
This expression is believed to define an appropriate starting
model for manganites, but unfortunately, it is quite difficult to
solve such a Hamiltonian.
In order to investigate further the properties of manganites,
some simplifications are needed.
In below, several types of simplified models are listed.

\subsubsection{One-orbital Model}

A simple model for manganites to illustrate the CMR effect is obtained
by neglecting the electron-phonon coupling and the Coulomb
interactions.
Usually, an extra simplification is carried out  by neglecting the
orbital degrees of freedom, leading to the FM Kondo model or
one-orbital double-exchange model, which will be simply referred as
the ``one-orbital model'' hereafter,
given as (Zener, 1951; Furukawa, 1994) 
\begin{eqnarray}
  H_{\rm DE} \!=\! -t \sum_{\langle{\bf i,j} \rangle,\sigma}
  (a_{{\bf i} \sigma}^{\dag}a_{{\bf j} \sigma}+{\rm H.c.})
  \!-\! J_{\rm H} \sum_{\bf i}{\bf s}_{\bf i} \cdot {\bf S}_{\bf j}
  \!+\! J_{\rm AF} \sum_{\langle {\bf i,j} \rangle}
  {\bf S}_{\bf i} \cdot {\bf S}_{\bf j},
\end{eqnarray}
where $a_{{\bf i} \sigma}$ is the annihilation operator for an
electron with spin $\sigma$ at site ${\bf i}$, but without orbital
index.
Note that $H_{\rm DE}$ is quadratic in the electron operators,
indicating that it is reduced to a one-electron problem on the
background of localized $t_{\rm 2g}$ spins.
This is a clear advantage for the Monte Carlo simulations.
Neglecting the orbital degrees of freedom is clearly an
oversimplification, and important phenomena such as orbital ordering
cannot be obtained in this model.
However, the one-orbital model is still important, since it already
includes part of the essence of manganese oxides.
For example, recent computational investigations have clarified that
the very important phase separation tendencies and metal-insulator
competition exist in this model.

\subsubsection{$J_{\rm H}$=$\infty$ limit}

Another simplification without the loss of essential physics is to
take the widely used limit $J_{\rm H}$=$\infty$,
since in the actual material $J_{\rm H}/t$ is much larger than unity.
In such a limit, the $e_{\rm g}$-electron spin perfectly aligns along 
the $t_{\rm 2g}$-spin direction, reducing the number of degrees of freedom.
Then, in order to diagonalize the Hund term,
the ``spinless"  $e_{\rm g}$-electron operator, $c_{{\bf i} \gamma}$,
is defined as
\begin{equation}
  c_{{\bf i} \gamma} =
  \cos(\theta_{\bf i}/2)d_{{\bf i}\gamma \uparrow}
  + \sin(\theta_{\bf i}/2)e^{-i\phi_{\bf i}}d_{{\bf i}\gamma\downarrow}.
\end{equation}
In terms of the $c$-variables, the kinetic energy acquires
the simpler form
\begin{equation}
H_{\rm kin} = -\sum_{{\bf ia}\gamma \gamma'}
  {\tilde t}^{\bf a}_{\gamma \gamma'} 
  c_{{\bf i} \gamma}^{\dag}c_{{\bf i+a} \gamma'},
\end{equation}
where ${\tilde t}^{\bf a}_{\gamma \gamma'}$ is defined as
${\tilde t}^{\bf a}_{\gamma \gamma'}$=
$S_{\bf i,i+a}t^{\bf a}_{\gamma \gamma'}$
with $S_{\bf i,j}$ given by
\begin{eqnarray}
  S_{\bf i,j} \!=\! \cos (\theta_{\bf i}/2)\cos (\theta_{\bf j}/2)
  \!+\! \sin (\theta_{\bf i}/2)\sin (\theta_{\bf j}/2)
  e^{-i(\phi_{\bf i}-\phi_{\bf j})}.
\end{eqnarray}
This factor denotes the change of hopping amplitude due to the
difference in angles between $t_{\rm 2g}$-spins at sites ${\bf i}$ and
${\bf j}$.
Note that the effective hopping in this case is a complex number
(Berry phase), contrary to the real number widely used in a large
number of previous investigations (for details in the case of the
one-orbital model see M\"uller-Hartmann and Dagotto, 1996).

The limit of infinite Hund coupling reduces the number of degrees of 
freedom substantially since the spin index is no longer needed.
In addition, the $U$- and $J$-terms in the electron-electron
interaction within the $e_{\rm g}$-sector are also no longer needed.
In this case, the following simplified model is obtained:
\begin{eqnarray}
  H^{\infty} \! &=& \!-\sum_{{\bf ia}\gamma \gamma'}
  {\tilde t}^{\bf a}_{\gamma \gamma'} 
  c_{{\bf i} \gamma}^{\dag}c_{{\bf i+a} \gamma'}
  \!+\! U' \sum_{\bf i} n_{{\bf i}{\rm a}}n_{{\bf i}{\rm b}}
  \!+\! V \sum_{\langle {\bf i,j} \rangle} n_{\bf i}n_{\bf j}
  \!+\! J_{\rm AF} \sum_{\langle {\bf i,j} \rangle}
  {\bf S}_{\bf i} \cdot {\bf S}_{\bf j}
  \nonumber \\ 
  &+& E_{\rm JT} \sum_{\bf i}
  [2( q_{1{\bf i}} n_{\bf i} + q_{2{\bf i}} \tau_{x{\bf i}}
  + q_{3{\bf i}} \tau_{z{\bf i}})
  + \beta q_{1{\bf i}}^2 + q_{2{\bf i}}^2 +q_{3{\bf i}}^2],
\end{eqnarray}
where 
$n_{{\bf i} \gamma}$=$c_{{\bf i} \gamma}^{\dag}c_{{\bf i} \gamma}$,
$n_{\bf i}$=$\sum_{\gamma}n_{{\bf i} \gamma}$,
$\tau_{x\bf i}$=
$c_{{\bf i}a}^{\dag}c_{{\bf i}b}$+$c_{{\bf i}b}^{\dag}c_{{\bf i}a}$,
and 
$\tau_{z\bf i}$=
$c_{{\bf i}a}^{\dag}c_{{\bf i}a}$$-$$c_{{\bf i}b}^{\dag}c_{{\bf i}b}$.

Considering the simplified Hamiltonian $H^{\infty}$, two other
limiting models can be obtained. One is the Jahn-Teller model 
$H^{\infty}_{\rm JT}$, defined as 
$H^{\infty}_{\rm JT}=H^{\infty}(U'=V=0)$,
in which the Coulomb interactions are simply ignored.
Another is the Coulombic model
$H^{\infty}_{\rm C}$, defined as 
$H^{\infty}_{\rm C}=H^{\infty}(E_{\rm JT}=0)$,
which denotes the two-orbital double exchange model 
influenced by the Coulomb interactions, neglecting the phonons.
Of course, the actual situation is characterized by 
$U' \ne 0$, $V \ne 0$, and $E_{\rm JT} \ne 0$,
but in the spirit of the adiabatic continuation,
it is convenient and quite meaningful to consider the 
minimal models possible to describe correctly the 
complicated properties of manganites.

\subsubsection{Jahn-Teller phononic and Coulombic models}

Another possible simplification could have been obtained by neglecting 
the electron-electron interaction in the full Hamiltonian
but keeping the Hund coupling finite, leading to the following purely
Jahn-Teller phononic model with active spin degrees of freedom:
\begin{equation}
  H_{\rm JT} = H_{\rm kin} + H_{\rm Hund} + H_{\rm AFM} 
  + H_{\rm el-ph}.
\end{equation}
Often in this chapter, this Hamiltonian will be referred to as the
``two-orbital'' model (unless explicitly stated otherwise).
To solve $H_{\rm JT}$, numerical methods such as Monte Carlo
techniques and the relaxation method have been employed.
Qualitatively, the negligible values of the probability of double
occupancy in the strong electron-phonon coupling region with large
$J_{\rm H}$ justifies the neglect of $H_{\rm el-el}$, since the
Jahn-Teller energy is maximized when one $e_{\rm g}$ electron exists
at each site. 
Thus, the Jahn-Teller phonon induced interaction will produce
physics quite similar to that due to the on-site correlation.

It would be important to verify this last expectation by studying a
multi-orbital model with only Coulombic terms, without the extra 
approximation of using mean-field techniques for its analysis.
Of particular relevance is whether phase separation tendencies and
charge ordering appear in this case, as they do in the Jahn-Teller
phononic model. 
This analysis is particularly important since, as explained before,
a mixture of phononic and Coulombic interactions is expected to be
needed for a proper quantitative description of manganites.
For this purpose, yet another simplified model has been analyzed
in the literature:
\begin{equation}
  H_{\rm C} = H_{\rm kin} + H_{\rm el-el}.
\end{equation}
Note that the Hund coupling term between $e_{\rm g}$ electrons and  
$t_{\rm 2g}$ spins is not explicitly included.
The reason for this extra simplification is that the numerical
complexity in the analysis of the model is drastically reduced by
neglecting the localized $t_{\rm 2g}$ spins.
In the FM phase, this is an excellent approximation, but not
necessarily for other magnetic arrangements. 
Nevertheless the authors believe that it is important to establish 
with accurate numerical techniques whether the phase separation tendencies
are already present in this simplified two-orbital models with Coulomb
interactions, even if not all degrees of freedom are incorporated from 
the outset.
Adding the $S$=3/2 quantum localized spins to the problem would
considerably increase the size of the Hilbert space of the model,
making it intractable with current computational techniques.

\subsection{Estimations of Parameters}

In this subsection, estimations of the couplings that appear in the
models described before are provided.
However, before proceeding with the details the reader must be warned
that such estimations are actually quite difficult, for the simple
reason that in order to compare experiments with theory reliable
calculations must be carried out.
Needless to say, strong coupling many-body problems are notoriously
difficult and complex, and it is quite hard to find accurate
calculations to compare against experiments.
Then, the numbers quoted below must be taken simply as rough
estimations of orders of magnitude. The reader should consult the
cited references to analyze the reliability of the estimations mentioned 
here.
Note also that the references discussed in this subsection correspond
to only a small fraction of the vast literature on the subject.
Nevertheless, the ``sample'' cited below is representative of the
currently accepted trends in manganites.

Regarding the largest energy scales, the on-site $U$ repulsion was
estimated to be 5.2$\pm$0.3 eV and 3.5$\pm$0.3 eV, for $\rm Ca Mn O_3$ 
and $\rm La Mn O_3$, respectively, by Park et al. (1996) using
photoemission techniques.
The charge-transfer energy has been found to be 3.0$\pm$0.5 eV for
$\rm Ca Mn O_3$ in the same study.
In other photoemission studies, Dessau and Shen (1999) estimated the
exchange energy for flipping an $e_{\rm g}$-electron to be 2.7eV.

Okimoto et al. (1995) studying the optical spectra of $\LSMO$ with
x=0.175 estimated the value of the Hund coupling to be of the order of 
2 eV, much larger than the hopping of the one-orbital model for
manganites.
Note that in estimations of this variety care must be taken with the
actual definition of the exchange $J_{\rm H}$, which sometimes is in
front of a ferromagnetic Heisenberg interaction where classical
localized spins of module 1 are used, while in other occasions quantum 
spins of value 3/2 are employed.  
Nevertheless, the main message of Okimoto et al.'s paper is that 
$J_{\rm H}$ is larger than the hopping.
A reanalysis of Okimoto et al.'s results led Millis, Mueller and
Shraiman (1996) to conclude that the Hund coupling is actually even
larger than previously believed.
The optical data of Quijada et al. (1998) and Machida et al. (1998) 
also suggest that the Hund coupling is larger than 1 eV.
Similar conclusions were reached by Satpathy et al. (1996) using
constrained LDA calculations.

The crystal-field splitting between the $e_{\rm g}$- and 
$t_{\rm 2g}$-states was estimated to be of the order of 1 eV by
Tokura (1999) (see also Yoshida, 1998). 
It is clear that manganites are in high-spin
ionic states due to their large Hund coupling.

Regarding the hopping ``$t$'', Dessau and Shen (1999) reported a value
of order 1eV, which is somewhat larger than other estimations.
In fact, the results of Bocquet et al. (1992), Arima et al. (1993),
and Saitoh et al. (1995) locate its magnitude between 
0.2 eV and 0.5 eV, which is reasonable in transition metal oxides.
However, note that fair comparisons between theory and experiment require
calculations of, e.g., quasiparticle band dispersions, which are
difficult at present. Nevertheless it is widely accepted that the
hopping is just a fraction of eV.

Dessau and Shen (1999) estimated the static Jahn-Teller energy 
$E_{\rm JT}$ as 0.25eV. From the static Jahn-Teller energy and the 
hopping amplitude, it is convenient to define the dimensionless
electron-phonon coupling constant $\lambda$ as
\begin{equation}
  \lambda=\sqrt{2E_{\rm JT}/t}=g/\sqrt{k_{\rm JT}t}.
\end{equation}
By using $E_{\rm JT}$=0.25eV and $t$=0.2$\sim$0.5eV,
$\lambda$ is estimated as between 1 $\sim$ 1.6.
Actually, Millis, Mueller and Shraiman (1996) concluded
that $\lambda$ is between 1.3 and 1.5.

As for the parameter $\beta$, it is given by 
$\beta$=$k_{\rm br}/k_{\rm JT}$=
$(\omega_{\rm br}/\omega_{\rm JT})^2$,
where $\omega_{\rm br}$ and $\omega_{\rm JT}$ are the vibration
energies for manganite breathing- and Jahn-Teller-modes, respectively, assuming
that the reduced masses for those modes are equal. From experimental 
results and band-calculation data (see Iliev et al. 1998),
$\omega_{\rm br}$ and $\omega_{\rm JT}$ are
estimated as $\sim 700$cm$^{-1}$ and $500$-$600$cm$^{-1}$,
respectively,
leading to $\beta$$\approx$2.
However, in practice it has been observed that the main conclusions
are basically unchanged as long as $\beta$ is larger than unity.
Thus, if an explicit value for $\beta$ is not provided, the reader can
consider that $\beta$ is simply taken to be $\infty$ to suppress the
breathing mode distortion.

The value of $J_{\rm AF}$ is the smallest of the set of couplings
discussed here. 
In units of the hopping, it is believed to be of the order of 0.1$t$ 
(see Perring et al., 1997), namely about 200K. Note, however, that it
would be a bad approximation to simply neglect this parameter since in
the limit of vanishing density of $e_{\rm g}$ electrons, $J_{\rm AF}$
is crucial to induce antiferromagnetism, as it occurs in 
$\rm Ca Mn O_3$ for instance.
Its relevance, at hole densities close to 0.5 or larger, to the
formation of antiferromagnetic charge-ordered states is remarked elsewhere
in this chapter.
Also in mean-field approximations by Maezono, Ishihara, and Nagaosa (1998)
the importance of $J_{\rm AF}$ has been mentioned, even though in their work
this coupling was estimated to be only 0.01$t$.

Summarizing, it appears well-established that:
(i) the largest energy scales in the Mn-oxide models studied here 
are the Coulomb repulsions between electrons in the same ion, which is 
quite reasonable.
(ii) The Hund coupling is between 1 and 2 eV, larger than the typical
hopping amplitudes, and sufficiently large to form high-spin Mn$^{4+}$
and Mn$^{3+}$ ionic states. 
As discussed elsewhere in this chapter, a large $J_{\rm H}$ leads
naturally to a vanishing probability of $e_{\rm g}$-electron
double-occupancy of a given orbital, thus mimicking the effect of a
strong on-site Coulomb repulsion. 
(iii) The dimensionless electron-phonon coupling constant $\lambda$
is of the order of unity, showing that the electron lattice
interaction is substantial and cannot be neglected.
(iv) The electron hopping energy is a fraction of eV.
(v) The AF-coupling among the localized spins is about a tenth of the
hopping. However, as remarked elsewhere, this apparent small coupling
can be quite important in the competition between FM and AF states.

\section{Spin-Charge-Orbital Ordering}

In the complicated phase diagram for manganites, there appear so many
magnetic phases.
A key concept to classify these phases is the charge and orbital ordering.
Especially, ``orbital ordering'' is the remarkable feature,
characteristic to manganites with active $e_{\rm g}$ orbital.
In this section, spin, charge, and orbital structure for the typical
hole doping in the phase diagram of manganites is focused by stressing
the importance of orbital ordering.

Note that the one-orbital model for manganites contains interesting
physics, notably a FM-AF competition that has similarities
with those found in experiments.
However, it is clear that to explain the notorious orbital order
tendency in Mn-oxides, it is crucial to use a model with two orbitals,
and in a previous section such a model was defined for the case where
there is an electron Jahn-Teller phonon coupling and also Coulomb
interactions.
Under the assumption that both localized $t_{\rm 2g}$ spins and
phonons are classical, the model without Coulombic terms can be
studied fairly accurately using numerical and mean-field
approximations.
For the case where Coulomb terms are included, unfortunately, 
computational studies
are difficult but mean-field approximations can still be carried out.

\begin{figure}[t]
\centerline{\epsfxsize=10.22truecm \epsfysize=7.84truecm \epsfbox{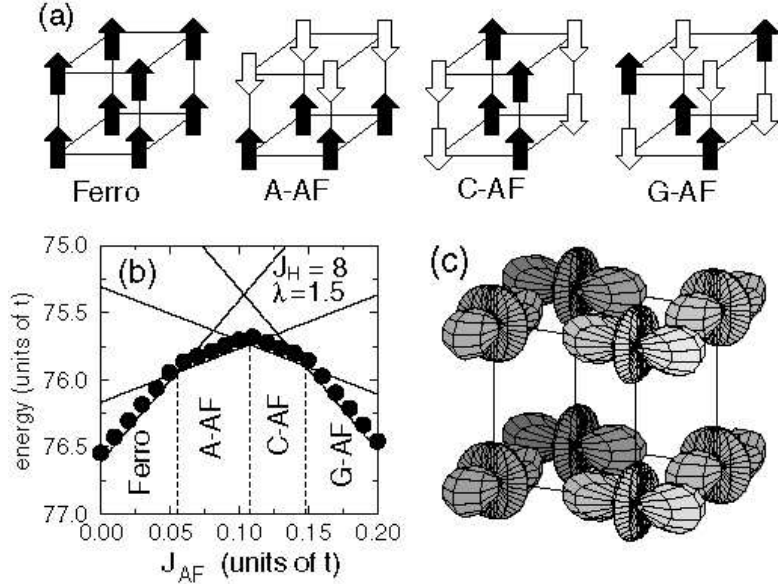}}
\caption{(a) The four spin arrangements for Ferro, A-AF, C-AF, and G-AF.
(b) Total energy vs $J_{\rm AF}$ on a 2$^3$ cluster at low temperature
with $J_{\rm H}$=8$t$ and $\lambda$=1.5. The results were obtained
using Monte Carlo and relaxational techniques, with excellent
agreement among them.
(c) Orbital order corresponding to the A-type AF state.
For more details the reader should consult Hotta et al., 1999.}
\label{fig3}
\end{figure}

\subsection{x=0.0}

First let us consider the mother material LaMnO$_3$ with one $e_{\rm g}$
electron per site.
This material has the insulating AF phase with  A-type AF spin order,
in which $t_{\rm 2g}$ spins are ferromagnetic in the $a$-$b$ plane and
antiferromagnetic along the $c$-axis.
Recent investigations by Hotta et al. (1999) have shown that, in the 
context of the model with Jahn-Teller phonons, the important ingredient
to understand the A-type AF phase is $J_{\rm AF}$,
namely by increasing this coupling from 0.05
to larger values, a transition from a FM to an A-type AF exists
(the relevance of Jahn-Teller couplings at \densi=1.0 has also been
remarked by Capone, Feinberg, and Grilli, 2000).
This can be visualized easily in Fig.~\ref{fig3},
where the energy vs. $J_{\rm AF}$ at fixed intermediate $\lambda$ and
$J_{\rm H}$ is shown.
Four regimes were identified: FM, A-AF, C-AF, and G-AF states that 
are sketched also in that figure.
The reason is simple: as $J_{\rm AF}$ grows, the tendency toward spin
AF must grow since this coupling favors such an order.
If $J_{\rm AF}$ is very large, then it is clear that a G-AF state must
be the one that lowers the energy, in agreement with the Monte Carlo 
simulations.
If $J_{\rm AF}$ is small or zero, there is no reason why spin AF will
be favorable at intermediate $\lambda$ and the density under consideration,
and then the state is ferromagnetic to improve the electronic mobility.
It should be no surprise that at intermediate $J_{\rm AF}$, the
dominant state is intermediate between the two extremes, with A-type
and C-type antiferromagnetism becoming stable in intermediate regions
of parameter space.

It is interesting to note that similar results regarding the relevance 
of $J_{\rm AF}$ to stabilize the A-type order have been found by
Koshibae et al. (1997) in a model with Coulomb interactions.
An analogous conclusion was found by 
Solovyev, Hamada, and Terakura (1996) and Ishihara et al. (1997).
Betouras and Fujimoto (1999), using bosonization techniques for the 
one-dimensional one-orbital model,
also emphasized the importance of $J_{\rm AF}$,
similarly as did Yi, Yu, and Lee (1999) based on Monte Carlo studies
in two dimensions of the same model.
The overall conclusion is that there are clear analogies between
the strong Coulomb and strong Jahn-Teller coupling approaches, as
discussed elsewhere in this chaper.
Actually in the mean-field approximation,
it was shown by Hotta, Malvezzi, and Dagotto (2000)
that the influence of the Coulombic terms
can be hidden in simple redefinitions of the electron-phonon couplings
(see also Benedetti and Zeyher, 1999).
In our opinion, both approaches (Jahn-Teller and Coulomb) have strong
similarities and it is not surprising that basically the same physics
is obtained in both cases.
Actually, Fig.~2 of Maezono, Ishihara, and Nagaosa (1998) showing
the energy vs. $J_{\rm AF}$ in mean-field calculations of the Coulombic
Hamiltonian without phonons is very similar to our Fig.~\ref{fig3},
aside from overall scales.
On the other hand, Mizokawa and Fujimori (1995, 1996) states that
the A-type AF is stabilized only when the Jahn-Teller distortion is
included, namely, the FM phase is stabilized in the purely Coulomb model,
based on the unrestricted Hartree-Fock calculation for the $d$-$p$ model.

The issue of what kind of orbital order is concomitant with A-type AF
order is an important matter. This has been discussed at length by 
Hotta et al. (1999), and the final conclusion, after the introduction
of perturbations caused by the experimentally known difference in
lattice spacings between the three axes, is that the order shown in
Fig.~\ref{fig3}(c) minimizes the energy. 
This state has indeed been identified in recent x-ray experiments, and
it is quite remarkable that such a complex pattern of spin and orbital
degrees of freedom indeed emerges from mean-field and computational
studies.
Studies by van den Brink et al. (1998) using purely Coulombic models
arrived at similar conclusions. 

Why does the orbital order occur here?
This can be easily understood perturbatively in the hopping $t$.
A hopping matrix only connecting the same orbitals,
with hopping parameter $t$, is assumed for simplicity. 
The energy difference between $e_{\rm g}$ orbitals at a given site is 
$E_{\rm JT}$, which is a monotonous function of $\lambda$.
For simplicity, in the notation let us refer to orbital uniform
(staggered) as orbital ``FM'' (``AF'').
Case (a) corresponds to spin FM and orbital AF: In this
case when an electron moves from orbital a on the left to the same
orbital on the right, which is the only possible hopping by assumption, 
an energy of order $E_{\rm JT}$ is lost, but kinetic energy is gained.
As in any second order perturbative calculation the energy gain is then
proportional to $t^2/E_{\rm JT}$.
In case (b), both spin and orbital FM, the electrons do not move and
the energy gain is zero (again, the nondiagonal hoppings are assumed
negligible just for simplicity).
In case (c), the spin are AF but the orbitals are FM.
This is like a one orbital model and the gain in energy is
proportional to $t^2/(2J_{\rm H})$.
Finally, in case (d) with AF in spin and orbital, both Hund and
orbital splitting energies are lost in the intermediate state, and the
overall gain becomes proportional to $t^2/(2J_{\rm H} + E_{\rm JT})$.
As a consequence, if the Hund coupling is larger than $E_{\rm JT}$,
then case (a) is the best, as it occurs at intermediate $E_{\rm JT}$ 
values.
Then, the presence of orbital order can be easily understood from a
perturbative estimation, quite similarly as done by Kugel and Khomskii (1974)
in their pioneering work on orbital order. 

Finally let us provide a comment on the FM tendency at x=0.0.
Carrying out Monte Carlo simulations in the localized spins and phonons,
and considering exactly the electrons in the absence of an explicit Coulomb
repulsion, a variety of correlations have been calculated
to establish the \densi=1.0 phase diagram.
Typical results for the spin and orbital structure factors, $S(q)$ and 
$T(q)$ respectively, were presented by Yunoki et al. (1998b)
At small $\lambda$, $S(0)$ is dominant and $T(q)$ is not active.
This signals a ferromagnetic state with disordered orbitals, namely a
standard ferromagnet
(note, however, that van den Brink and Khomskii (2001)
and Maezono and Nagaosa (2000) believe that this state in experiments
may have complex orbital ordering).  
The result with FM tendencies dominating may naively seem strange
given the fact that for the one-orbital model at \densi=1.0 an AF
state was found. 
But here two orbitals are being considered and one electron per site
is 1/2 electron per orbital.
In this respect, \densi=1.0 with two orbitals should be similar to 
\densi=0.5 for one orbital and indeed in the last case a ferromagnetic 
state was observed.


\begin{figure}[t]
\centerline{\epsfxsize=9.truecm \epsfbox{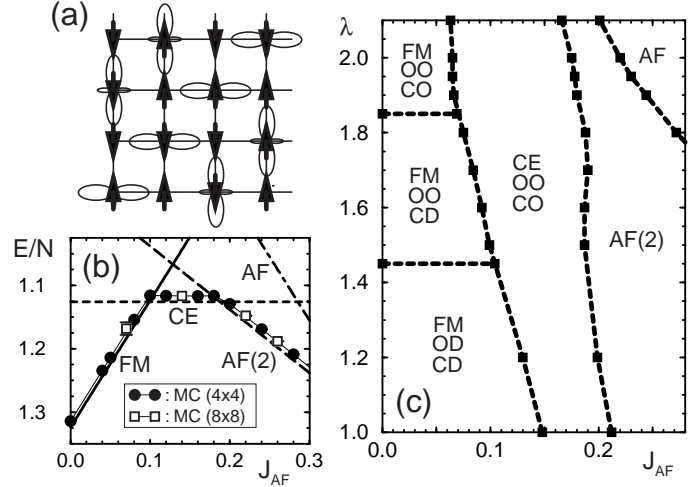}}
\caption{(a) Schematic view of CE-type structure at x=0.5.
(b) Monte Carlo energy per site vs $J_{\rm AF}$ at density x=0.5,
$\lambda$=1.5, low temperature $T$=1/100, and $J_{\rm H}$=$\infty$,
using the two-orbital model in two dimensions with Jahn-Teller phonons
(non-cooperative ones).
FM, CE, and AF states were identified measuring charge, spin, and 
orbital correlations. ``AF(2)'' denotes a state with
spins $\uparrow \uparrow \downarrow \downarrow$ in one direction,
and antiferromagnetically coupled in the other. The clusters used are 
indicated.
(c) Phase diagram in the plane $\lambda$-$J_{\rm AF}$ at x=0.5,
obtained numerically using up to 8$\times$8 clusters. All transitions
are of first-order. The notation is the standard one (CD = charge
disorder, CO = charge order, OO = orbital order, OD = orbital
disorder). Results reproduced from Yunoki, Hotta and Dagotto (2000),
where more details can be found.}
\label{fig4}
\end{figure}

\subsection{x=0.5}

Now let us move to another important doping x=0.5.
For half-doped perovskite manganites, the so-called CE-type AFM phase
has been established as the ground state in the 1950's. 
This phase is composed of zigzag FM arrays of $t_{\rm 2g}$-spins, 
which are coupled antiferromagnetically
perpendicular to the zigzag direction.
Furthermore, the checkerboard-type charge ordering in the $x$-$y$
plane, the charge stacking along the $z$-axis, and
($3x^2-r^2$/$3y^2-r^2$) orbital ordering are associated with this phase.
A schematic view of CE-type structure is shown in Fig.~\ref{fig4}(a).

Although there is little doubt that the famous CE-state of Goodenough
is indeed the ground state of x=0.5
intermediate and low bandwidth manganites, only very recently such a
state has received theoretical confirmation using unbiased techniques, 
at least within some models.
In the early approach of Goodenough it was $assumed$ that the charge
was distributed in a checkerboard pattern, upon which spin and orbital
order was found. But it would be desirable to obtain the CE-state
based entirely upon a more fundamental theoretical analysis, as
the true state of minimum energy of a well-defined and realistic
Hamiltonian.
If such a calculation can be done, as a bonus one would find out which 
states compete with the CE-state in parameter space, an
issue very important in view of the mixed-phase tendencies of
Mn-oxides, which cannot be handled within the approach of Goodenough.

One may naively believe that it is as easy as introducing a huge
nearest-neighbor Coulomb repulsion $V$ to stabilize a charge-ordered
state at x=0.5, upon which the reasoning of Goodenough can be applied.
However, there are at least two problems with this approach.
First, such a large $V$ quite likely will destabilize the
ferromagnetic charge-disordered state and others supposed to be
competing with the CE-state. It may be possible to explain the
CE-state with this approach, but not others also observed at
x=0.5 in large bandwidth Mn-oxides.
Second, a large $V$ would produce a checkerboard pattern in the
$three$ directions.
However, experimentally it has been known for a long time (Wollan and
Koehler, 1955) that the charge $stacks$ along the $z$-axis, namely the
same checkerboard pattern is repeated along $z$, rather than being
shifted by one lattice spacing from plane to plane.
A dominant Coulomb interaction $V$ can not be the whole story
for x=0.5 low-bandwidth manganese oxides.

The nontrivial task of finding a CE-state with charge stacked along
the $z$-axis without the use of a huge nearest-neighbors repulsion
has been recently performed by Yunoki, Hotta, and Dagotto (2000) 
using the two-orbital model with strong electron Jahn-Teller phonon
coupling.
The calculation proceeded using an unbiased Monte Carlo simulation,
and as an output of the study, the CE-state indeed emerged as the
ground-state in some region of coupling space.
Typical results are shown in Figs.~\ref{fig4}(b) and (c).
In part (b) the energy at very low temperature is shown as a function
of $J_{\rm AF}$ at fixed density x=0.5, $J_{\rm H}$=$\infty$ for
simplicity, and with a robust electron-phonon coupling $\lambda$=1.5
using the two orbital model $H_{\rm JT}$
At small $J_{\rm AF}$, a ferromagnetic phase was found to be
stabilized, according to the Monte Carlo simulation.
Actually, at $J_{\rm AF}$=0.0 it has not been possible to stabilize a
partially AF-state at x=0.5, namely the states are always
ferromagnetic at least within the wide range of $\lambda$'s
investigated (but they can have charge and orbital order).
On the other hand, as $J_{\rm AF}$ grows, a tendency to form AF links
develops, as it happens at x=0.0.
At large $J_{\rm AF}$ eventually the system transitions to
states that are mostly antiferromagnetic, such as the so-called
``AF(2)'' state of Fig.~\ref{fig4}(b) (with an up-up-down-down spin pattern
repeated along one axis, and AF coupling along the other axis),
or directly a fully AF-state in both directions. 

However, the intermediate values of $J_{\rm AF}$ are the most
interesting ones. In this case the energy of the two-dimensional clusters 
become flat as a function of $J_{\rm AF}$ 
suggesting that the state has the same
number of FM and AF links, a property that the CE-state indeed has.
By measuring charge-correlations it was found that a checkerboard
pattern is formed particularly at intermediate and large $\lambda$'s,
as in the CE-state.
Finally, after measuring the spin and orbital correlations, it was
confirmed that indeed the complex pattern of the CE-state was fully
stabilized in the simulation. This occurs in a robust portion of the
$\lambda$-$J_{\rm AF}$ plane, as shown in Fig.~\ref{fig4}(c).
The use of $J_{\rm AF}$ as the natural parameter to vary in order
to understand the CE-state is justified based on Fig.~\ref{fig4},
since the region of stability of the CE-phase is elongated
along the $\lambda$-axis,
meaning that its existence is not so much dependent on that coupling
but much more on $J_{\rm AF}$ itself.
It appears that some explicit tendency in the Hamiltonian toward the
formation of AF links is necessary to form the CE-state.
If this tendency is absent, a FM state if formed, while if it is too 
strong an AF-state appears. The x=0.5 CE-state, similar to the 
A-type AF at x=0.0, needs an intermediate value of $J_{\rm AF}$ for
stabilization. The stability window is finite and in this respect there
is no need to carry out a $fine$ tuning of parameters to find the CE
phase. However, it is clear that there is a balance of AF and FM
tendencies in the CE-phase that makes the state somewhat fragile.

Note that the transitions among the many states obtained when varying
$J_{\rm AF}$ are all of $first$ order, namely they correspond to
crossings of levels at zero temperature.
The first-order character of these transitions is a crucial ingredient
of the recent scenario proposed by Moreo et al. (2000) involving
mixed-phase tendencies with coexisting clusters with $equal$ density.
Recently, first-order transitions have also been reported in the
one-orbital model at x=0.5 by Alonso et al. (2001a, 2001b), as well as
tendencies toward phase separation. Recent progress in the development
of powerful techniques for manganite models (Alonso et al., 2001c;
Motome and Furukawa, 1999, 2000) will
contribute to the clarification of these issues in the near future.

\begin{figure}[t]
\centerline{\epsfxsize=11.truecm \epsfbox{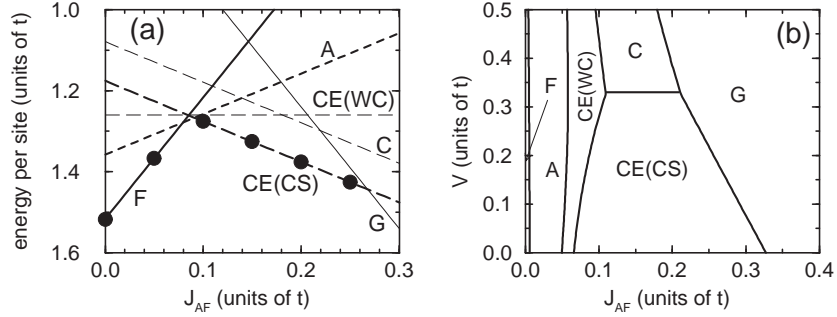}}
\caption{(a) Energy per site as a function of $J_{\rm AF}$ for
$\lambda$=1.6 and $J_{\rm H}$=$\infty$ for $H_{\rm JT}$. 
The curves denote the mean-field results and the solid symbols
indicate the energy obtained by the relaxation method.
Thick solid, thick broken, thin broken, thick dashed, thin dashed,
thin broken, and thin solid lines denotes FM, A-type, CE-type
with WC structure, charge-stacked CE-type, C-type, and G-type states,
respectively.
Note that the charge-stacked CE-state is observed in experiments.
(d) Phase diagram in the $(J_{\rm AF},V)$ plane. Note that the
charge-stacked structure along the $z$-axis can be observed 
only in the CE-type AFM phase.
Results reproduced from Hotta, Malvezzi, and Dagotto (2000),
where more details can be found.}
\label{fig5}
\end{figure}

Let us address now the issue of charge-stacking (CS) along the $z$-axis.
For this purpose simulations using three-dimensional
clusters were carried out.
The result for the energy vs. $J_{\rm AF}$ is shown in Fig.~\ref{fig5}(a),
with $J_{\rm H}$=$\infty$ and $\lambda$=1.6 fixed.
The CE-state with charge-stacking has been found to be the ground state 
on a wide $J_{\rm AF}$ window.
The reason that this state has lower energy than the so-called
``Wigner-crystal'' (WC) version of the CE-state, namely with the
charge spread as much as possible, is once again the influence of 
$J_{\rm AF}$. With a charge stacked arrangement, the links along the
$z$-axis can all be simultaneously antiferromagnetic, thereby minimizing
the energy. In the WC-state this is not possible.

It should be noted that this charge stacked CE-state is not
immediately destroyed when the weak nearest-neighbor repulsion $V$
is introduced to the model, as shown in Fig.~\ref{fig5}(b),
obtained in the mean-field calculations by
Hotta, Malvezzi, and Dagotto (2000).
If $V$ is further increased for a realistic value of $J_{\rm AF}$,
the ground state eventually changes from the charge stacked
CE-phase to the WC version of the CE-state or the C-type AFM phase 
with WC charge ordering.
As explained above, the stability of the charge stacked phase to 
the WC version of the CE-state is due to the magnetic energy difference.
However, the competition between the charge-stacked CE-state
and the C-type AFM phase with the WC structure is not simply understood 
by the effect of $J_{\rm AF}$, since those two kinds of AFM phases
have the same magnetic energy.
In this case, the stabilization of the charge stacking originates
from the difference in the geometry of the one-dimensional FM path,
namely a zigzag-path for the CE-phase and a straight-line path
for the C-type AFM state.
As will be discussed later, the energy for $e_{\rm g}$
electrons in the zigzag path is lower than that in the straight-line
path, and this energy difference causes the stabilization of the
charge stacking.
In short, the stability of the charge-stacked structure
at the expense of $V$ is supported by ``the geometric energy''
as well as the magnetic energy.
Note that each energy gain is just a fraction of $t$.
Thus, in the absence of other mechanisms to understand the
charge-stacking, another consequence of this analysis is that $V$
actually must be substantially $smaller$ than naively expected,
otherwise such a charge pattern would not be stable.
In fact, estimations given by Yunoki, Hotta, and Dagotto (2000)
suggest that the manganites must have a large dielectric function at
short distances (see Arima and Tokura, 1995) to prevent the melting
of the charge-stacked state.

Note also that the mean-field approximations by Hotta, Malvezzi, and
Dagotto (2000) have shown that on-site Coulomb interactions $U$ and
$U'$ can $also$ generate a two-dimensional CE-state, in agreement with the
calculations by van den Brink et al. (1999).
Then, the present authors believe that strong Jahn-Teller and Coulomb couplings 
tend to give similar results.
This belief finds partial confirmation in the mean-field
approximations of Hotta, Malvezzi, and Dagotto (2000), where the
similarities between a strong $\lambda$ and $(U,U')$ were
investigated.
Even doing the calculation with Coulombic interactions, the influence 
of $J_{\rm AF}$ is still crucial to inducing charge-stacking (note that 
the importance of this parameter has also been recently remarked by
Mathieu, Svedlindh and Nordblad, 2000, based on experimental results).

Many other authors carried out important work in the context of the
CE-state at x=0.5.
For example, with the help of Hartree-Fock calculations, Mizokawa and
Fujimori (1997) reported the stabilization of the CE-state at x=0.5
only if Jahn-Teller distortions were incorporated into a model with
Coulomb interactions.
This state was found to be in competition with a uniform FM state, as
well as with an A-type AF-state with uniform orbital order.
In this respect the results are very similar to those found by Yunoki,
Hotta and Dagotto (2000) using Monte Carlo simulations.
In addition, using a large nearest neighbor repulsion and the
one-orbital model, charge ordering and a spin structure compatible
with the zigzag chains of the CE state was found by Lee and Min (1997)
at x=0.5.
Also Jackeli et al. (1999) obtained charge-ordering at x=0.5 using
mean-field approximations and a large $V$.
Charge-stacking was not investigated by those authors.
The CE-state in x=0.5 $\PCMO$ was also obtained
by Anisimov et al. (1997) using LSDA+U techniques.

\subsection{x$>$0.5}

In the previous subsection, the discussion focused on the CE-type AFM 
phase at x=0.5.
Naively, it may be expected that similar arguments can be extended to
the regime x$>$1/2, since in the phase diagram for $\LCMO$, the AFM
phase has been found at low temperatures in the region
0.50$<$x$<$0.88.
Then, let us try to consider the band-insulating phase for density
x=2/3 based on $H^{\infty}$, without both the Jahn-Teller phononic and
Coulombic interactions, since this doping is quite important for the
appearance of the bi-stripe structure (see Mori et al., 1998).

After several calculations for x=2/3, as reported by Hotta et al.
(2000), the lowest-energy state was found to be characterized by the
straight path, not the zigzag one, leading to the C-type AFM phase
which was also discussed in previous Sections.
For a visual representation of the C-type state,
see Fig.~4 of Kajimoto et al., 1999.
At first glance, the zigzag structure similar to that for x=0.5
could be the ground-state for the same reason
as it occurs in the case of x=0.5. 
However, while it is true that the state with such a zigzag structure
is a band-insulator, the energy gain due to the opening of the bandgap
is not always the dominant effect.
In fact, even in the case of x=0.5, the energy of the bottom of the
band for the straight path is $-2t_0$, while for the zigzag path,
it is $-\sqrt{3}t_0$. For x=1/2, the energy gain due to the gap
opening overcomes the energy difference at the bottom of the band,
leading to the band-insulating ground-state. 
However, for x=2/3 even if a band-gap opens the energy of the zigzag
structure cannot be lower than that of the metallic straight-line
phase. Intuitively, this point can be understood as follows: 
An electron can move smoothly along the one-dimensional path
if it is straight. However, if the path is zigzag, ``reflection" of
the wavefunction occurs at the corner, and then a smooth movement of
one electron is no longer possible. Thus, for small numbers of
carriers, it is natural that the ground-state is characterized by the
straight path to optimize the kinetic energy of the $e_{\rm g}$
electrons. 

However, in neutron scattering experiments a spin pattern similar 
to the CE-type AFM phase has been suggested (Radaelli et al., 1999).
In order to stabilize the zigzag AFM phase to reproduce those
experiments it is necessary to include the Jahn-Teller distortion effectively. 
As discussed in Hotta et al. (2000), a variety of zigzag paths could
be stabilized when the Jahn-Teller phonons are included.
In such a case, the classification of zigzag paths is an important
issue to understand the competing ``bi-stripe" vs. ``Wigner-crystal"
structures.
The former has been proposed by Mori et al. (1998), while the latter 
was claimed to be stable by Radaelli et al. (1999).
In the scenario by Hotta et al.(2000), the shape of the zigzag structure is
characterized by the ``winding number'' $w$ associated with the 
Berry-phase connection of an $e_{\rm g}$-electron parallel-transported 
through Jahn-Teller centers, along zigzag one-dimensional paths
Namely, it is defined as
\begin{equation}
  \label{winding}
  w= \oint {d{\bf r} \over 2\pi} \nabla \xi.
\end{equation}
This quantity has been proven to be an integer, which is a topological
invariant (See Hotta et al., 1998.
See also Koizumi et al., 1998a and 1998b). 
Note that the integral indicates an accumulation of the phase
difference along the one-dimensional FM path in the unit length.
This quantity is equal to half of the number of corners included in
the unit path, which can be shown as follows. 
The orbital polarizes along the hopping direction, indicating that
$\xi_{\bf i}$=$2\pi/3$($4\pi/3$) along the 
$x$-($y$-)direction, as was pointed out above.
This is simply the double exchange mechanism in the orbital degree of
freedom.
Thus, the phase does not change in the straight segment part,
indicating that $w$=0 for the straight-line path. 
However, when an $e_{\rm g}$-electron passes a corner site, the
hopping direction is changed, indicating that the phase change occurs
at that corner.
When the same $e_{\rm g}$-electron passes the next corner, the hopping
direction is again changed.
Then, the phase change in $\xi_{\bf i}$ after moving through a couple
of corners should be $2\pi$, leading to an increase of unity in $w$. 
Thus, the total winding number is equal to half of the number of corners
included in the zigzag unit path. 
Namely, the winding number $w$ is a good label to specify the shape of
the zigzag one-dimensional FM path.

\begin{figure}[t]
\centerline{\epsfxsize=11.truecm \epsfbox{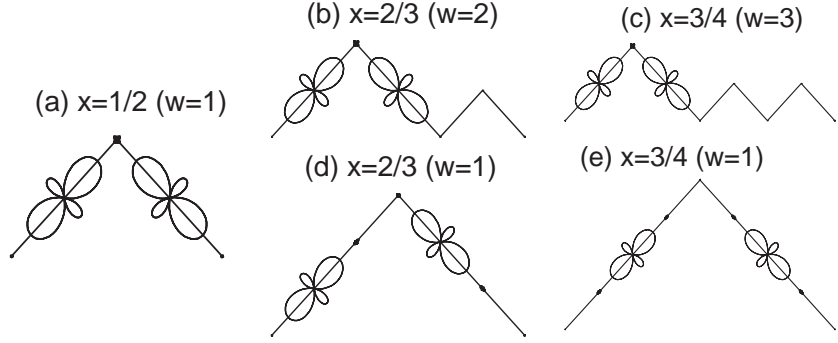}}
\caption{(a) Path with $w$=$1$ at x=$1/2$.
Charge and orbital densities are
calculated in the MFA for $E_{\rm JT}$=$2t$. At each site, the orbital
shape is shown with its size in proportion to the orbital density.
(b) The BS-structure path with $w$=$2$ at x=$2/3$. 
(c) The BS-structure path with $w$=$3$ at x=$3/4$. 
(d) The WC-structure path with $w$=$1$ at x=$2/3$. 
(e) The WC-structure path with $w$=$1$ at x=$3/4$.}
\label{fig6}
\end{figure}

\begin{figure}[t]
\centerline{\epsfxsize=11.truecm \epsfbox{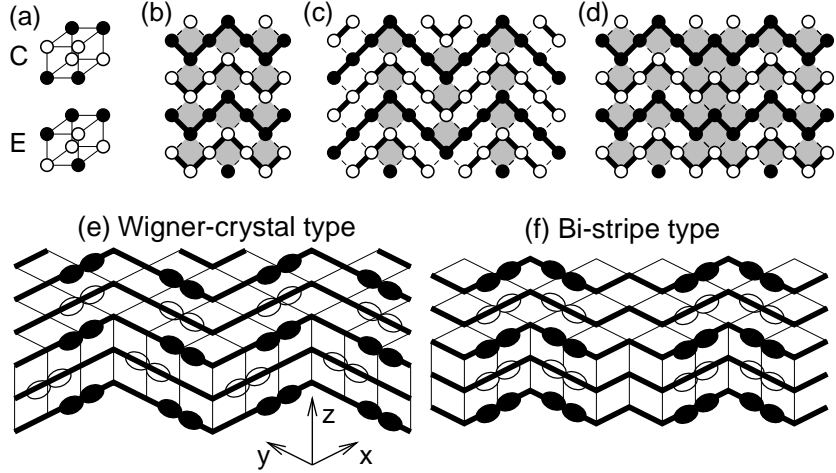}}
\caption{(a) C- and E-type unit cell (Wollan and Koehler, 1955).
(b) The spin structure in the $a$-$b$ plane at x=1/2.
Open and solid circle denote the spin up and down, respectively.
The thick line indicates the zigzag FM path.
The open and shaded squares denote the C- and E-type 
unit cells.
At x=1/2, C-type unit cell occupies half of the two-dimensional plane,
clearly indicating the ``CE'' type phase.
(c) The spin structure at x=2/3 for Wigner-crystal type
phase. Note that 66\% of the two-dimensional lattice is occupied
by C-type unit cell. Thus, it is called ``C$_2$E''-type AFM phase.
(d) The spin structure at x=2/3 for bi-stripe type
phase. Note that 33\% of the two-dimensional lattice is occupied
by C-type unit cell. Thus, it is called ``CE$_2$''-type AFM phase.
Schematic figures for spin, charge, and orbital ordering for (e) WC
and (f) BS structures at x=2/3. The open and solid symbols indicate
the spin up and down, respectively. The FM one-dimenaional path is
denoted by the thick line.
The empty sites denote Mn$^{4+}$ ions, while the robes
indicate the Mn$^{3+}$ ions in which $3x^2-r^2$ or $3y^2-r^2$ orbitals
are occupied.}
\label{fig7}
\end{figure}

After several attempts to include effectively the Jahn-Teller phonons, it was
found that the bi-stripe phase and the Wigner crystal phase
universally appear for $w$=$\rm x/(1-x)$ and $w$=1, respectively.
Note here that the winding number for the bi-stripe structure has a
remarkable dependence on x, reflecting  the fact that the distance
between adjacent bi-stripes changes with x.
This x-dependence of the modulation vector of the lattice distortion
has been observed in electron microscopy experiments 
(Mori et al., 1998).
The corresponding zigzag paths with the charge and orbital ordering 
are shown in Fig.~\ref{fig6}. In the bi-stripe structure, the charge is
confined in the short straight segment as in the case of the CE-type
structure at x=0.5.
On the other hand, in the Wigner-crystal structure, the straight
segment includes two sites, indicating that the charge prefers to
occupy either of these sites. 
Then, to minimize the Jahn-Teller energy and/or the Coulomb repulsion, the
$e_{\rm g}$ electrons are distributed with equal spacing. 
The corresponding spin structure is shown in Fig.~\ref{fig7}.
A difference in the zigzag geometry can produce a significant
different in the spin structure.
Following the definitions for the C- and E-type AFM structures
(see Wollan and Koehler, 1955),
the bi-stripe and Wigner crystal structure have $\rm C_{1-x}E_{x}$-type
and $\rm C_{x}E_{1-x}$-type AFM spin arrangements, respectively.
Note that at x=1/2, half of the plane is filled by the C-type,
while another half is covered by the E-type, clearly illustrating the
meaning of ``CE" in the spin structure of half-doped manganites.

The charge structure along the $z$-axis for x=2/3 has been discussed by
Hotta et al. (2000), as schematically shown in Figs.~\ref{fig7}(e) and (f),
a remarkable feature can be observed.
Due to the confinement of charge in the short straight segment for the
bi-stripe phase, the charge stacking is suggested from our topological
argument. 
On the other hand, in the Wigner-crystal type structure, charge is not
stacked, but it is shifted by one lattice constant to avoid the
Coulomb repulsion. 
Thus, if the charge stacking is also observed in the experiment for
x=2/3, our topological scenario suggests the bi-stripe phase as the
ground-state in the low temperature region.
To firmly establish the final ``winner" in the competition between the
bi-stripe and Wigner-crystal structure at x=2/3, more precise
experiments, as well as quantitative calculations, will be needed in the
future. 

\subsection{x$<$0.5}

Regarding densities smaller than 0.5, the states at x=1/8, 1/4 and
3/8 have received considerable attention recently (see Mizokawa et
al., 2000; Korotin et al., 1999; Hotta and Dagotto, 2000).
These investigations are still in a ``fluid'' state, and the
experiments are not quite decisive yet, and for this reason, 
this issue will not be discussed in much detail here.
However, without a doubt, it is very important to clarify the
structure of charge-ordered states that may be in competition  
with the ferromagnetic states in the range in which the latter is
stable in some compounds.
``Stripes'' may emerge from this picture, as recently remarked in
experiments
(Adams et al., 2000; Dai et al., 2000; Kubota et al., 2000.
See also Vasiliu-Doloc et al., 1999) and 
calculations (Hotta, Feiguin, and Dagotto, 2000), and surely the
identification of charge/orbital arrangements at x$<$0.5 will be an
important area of investigations in the very near future.

\begin{figure}[t]
\centerline{\epsfxsize=10.68truecm \epsfysize=8.76truecm \epsfbox{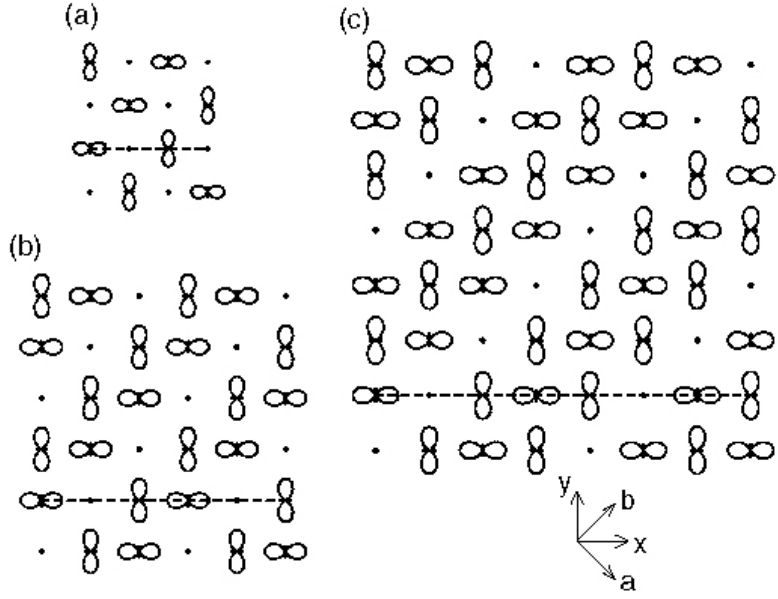}}
\caption{Orbital densities in the FM phase
for (a)x=1/2, (b)1/3, and (c)1/4.
The charge density in the lower-energy orbital is shown, and the size
of the orbital is in proportion to this density. The broken line
indicates one of the periodic paths to cover the whole two-dimensional plane.}
\label{fig8}
\end{figure}

Here a typical result for this stripe-like charge ordering is shown in
Fig.~\ref{fig8}, in which the lower-energy orbital at each site is depicted,
and its size is in proportion to the electron density occupying that
orbital.
This pattern is theoretically obtained by the relaxation technique for
the optimization of oxygen positions, namely including the cooperative
Jahn-Teller effect. 
At least in the strong electron-phonon coupling region, the stripe
charge ordering along the $diagonal$ direction in the $x$-$y$ plane
becomes the global ground-state.
Note, however, that many meta-stable states can appear very close to
this ground state. 
Thus, the shape of the stripe is considered to fluctuate both in space 
and time, and in experiments it may occur that only some fragments of
this stripe can be detected. 
It should also be emphasized that the orbital ordering occurs concomitant
with this stripe charge ordering.  
In the electron-rich region, the same antiferro orbital-order exists
as that corresponding to x=0.0.
On the other hand, the pattern around the diagonal array of
electron-poor sites is quite similar to the building block of the
charge/orbital structure at x=0.5. 

If these figures are rotated by 45 degrees,
the same charge and orbital structure is found to stack
along the $b$-axis.
Namely, it is possible to cover the whole two-dimensional plane by 
some periodic charge-orbital array along the $a$-axis
(see, for instance, the broken-line path).
If this periodic array is taken as the closed loop $C$ in 
Eq.~(\ref{winding}), the winding numbers are $w$=1, 2, and 3,
for x=1/2, 1/3, and 1/4, respectively.
Note that in this case $w$ is independent of the path along the $a$-axis.
The results imply a general relation $w$=$(1-x)/x$
for the charge-orbital stripe in the FM phase, reflecting the fact that
the distance between the diagonal arrays of holes changes with x.
Our topological argument predicts stable charge-orbital stripes at special
doping such as x=$1/(1+w)$, with $w$ an integer.

This orbital ordering can be also interpreted as providing  a ``$\pi$"-shift
in the orbital sector, by analogy with the dynamical stripes found in
cuprates (see, for instance, Buhler et al., 2000), although in copper oxides
the charge/spin stripes mainly appear along the $x$- or $y$-directions.
The study of the similarities and differences between stripes in manganites
and cuprates is one of the most interesting open problems in the study of
transition metal oxides, and considerable work is expected in the near future.

\begin{figure}[t]
\centerline{\epsfxsize=6.truecm \epsfbox{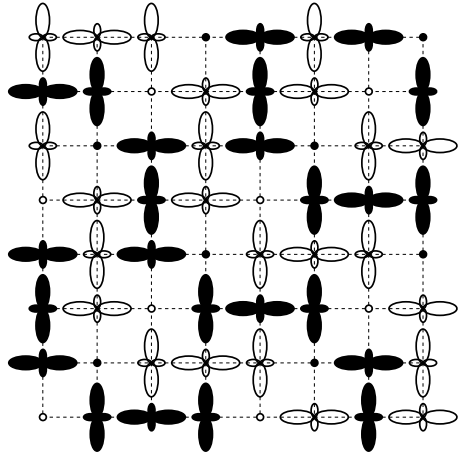}}
\caption{Schematic figure of the spin-charge-orbital structure
at x=1/4 in the zigzag AFM phase at low temperature and large electron-phonon
coupling. The symbol convention is the same as in Fig.~1.7.
This figure was obtained using numerical techniques,
and $cooperative$ phonons, for $J_{\rm H}$=$\infty$ and $J_{\rm AF}$=$0.1t$.
For the non-cooperative phonons, basically the same pattern
can be obtained.}
\label{fig9}
\end{figure}

Finally, a new zigzag AFM spin configuration for x$<$0.5 is here briefly
discussed (Hotta, Feiguin, and Dagotto, 2000).
In Fig.~\ref{fig9},
a schematic view of this novel spin-charge-orbital structure
on the 8$\times$8 lattice at x=1/4 is shown, deduced using the numerical
relaxation technique applied to cooperative Jahn-Teller phonons
in the strong-coupling region.
This structure appears to be the global ground state, but many excited states
with different spin and charge structures are also found with small
excitation energy, suggesting that the AFM spin structure for x$<$0.5
in the layered manganites is easily disordered due to this
``quasi-degeneracy'' in the ground state.
This result may be related to the ``spin-glass'' nature of the single layer
manganites reported in experiments (see Moritomo et al., 1995).

It should be noted that the charge-orbital structure is essentially the
same as that in the two-dimensional FM phase (see Fig.~\ref{fig9}).
This suggests the following scenario for the layered manganites:
When the temperature is decreased from the higher temperature region,
first charge ordering occurs due to the cooperative Jahn-Teller
distortions in the FM (or paramagnetic) region.
If the temperature is further decreased, the zigzag AFM spin arrangement
is stabilized, adjusting itself to the orbital structure.
Thus, the separation between the charge ordering temperature $T_{\rm CO}$
and the N\'eel temperature $T_{\rm N}$ occurs naturally in this context.
This is not surprising, since $T_{\rm CO}$ is due to the electron-lattice 
coupling, while $T_{\rm N}$ originates in the coupling $J_{\rm AF}$.
However, if the electron-phonon coupling is weak, then $T_{\rm CO}$ becomes
very low. In this case, the transition to the zigzag AFM phase may occur
prior to the charge ordering. 
As discussed above, the $e_{\rm g}$ electron hopping is confined to
one dimensional structures in the zigzag AFM environment.
Thus, in this situation, even a weak coupling electron-phonon coupling can
produce the charge-orbital ordering, as easily understood from the Peierls
instability argument.
Namely, just at the transition to the zigzag AFM phase, the charge-orbital
ordering occurs simultaneously, indicating that $T_{\rm CO}$=$T_{\rm N}$.
Note also that in the zigzag AFM phase, there is no essential difference
in the charge-orbital structures for the non-cooperative and cooperative
phonons, due to the one-dimensionality of those zigzag chains.

\subsection{Orbital ordering in related material}

In previous sections the charge and orbital ordering has been discussed
in detail for manganites, but recently it has been widely recognized that
orbital ordering is ubiquitous in transition metal oxides and
f-electron systems.
Among them an importance of hidden orbital ordering has been suggested
in the single-layered ruthenate by Hotta and Dagotto (2001).
As is well known, Sr$_2$RuO$_4$ is triplet superconductor (see Maeno,
Rice, and Sigrist, 2001).
When Sr is partially substituted by Ca, superconductivity is rapidly
destroyed and a paramagnetic metallic phase appears, while for
Ca$_{1.5}$Sr$_{0.5}$RuO$_4$, a nearly FM metallic phase has been suggested.
Upon further substitution, the system eventually transforms into
an AFM insulator (Nakatsuji and Maeno, 2000).
The G-type AFM phase in $\rm Ca_2RuO_4$ is characterized as a standard N\'eel
state with spin $S$=1 (Nakatsuji et al., 1997 and Braden et al., 1998).
To understand the N\'eel state observed in experiments,
one may consider the effect of the tetragonal
crystal field, leading to the splitting between xy and \{yz,zx\}
orbitals, where the xy-orbital state is lower in energy than the other levels.
When the xy-orbital is fully occupied, 
a simple superexchange interaction at strong Hund coupling
can stabilize the AFM state.
However, recent X-ray absorption spectroscopy studies have shown that
0.5 holes per site exist in the xy-orbital, while 1.5 holes are contained
in the zx- and yz-orbitals (Mizokawa et al., 2001), suggesting that the above
naive picture based on crystal field effects is incomplete.
This fact suggests that the orbital degree of freedom may play
a more crucial role in the magnetic ordering in ruthenates than
previously anticipated.

\begin{figure}[t]
\centerline{\epsfxsize=4.truecm \epsfbox{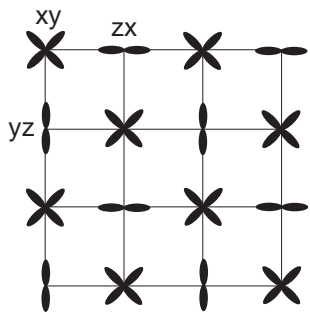}}
\caption{Schematic view of orbital ordering pattern proposed for
the G-type AFM phase of Ca$_2$RuO$_4$.}
\label{fig10}
\end{figure}

In order to understand the G-AFM phase with peculiar hole
arrangement, the 3-orbital ($t_{\rm 2g}$) Hubbard model tightly coupled
to lattice distortions has been analyzed using numerical and mean-field 
techniques (Hotta and Dagotto, 2001).
Note here that Ru$^{4+}$ ion takes the low spin state, in which four electrons
occupy $t_{\rm 2g}$ orbital.
Since this model includes several degrees of freedom, namely 2-spin and
3-orbital per electron,
it is quite difficult to study large-size clusters.
However, it is believed that the essential character of the competing states
can be captured using a small cluster, through the combination of the Lanczos
method and relaxational techniques.
Mean-field approximations complement and support the results obtained
numerically. An important conclusion of this analysis
is that the G-type AFM phase is stabilized only
when {\it both} Coulombic and phononic interactions are taken into account.
The existence of a novel orbital ordering (see Fig.~\ref{fig10}) is crucial
to reproduce the peculiar hole arrangement observed in 
experiments by Mizokawa et al., 2001.
Another interesting consequence of this study is the possibility
of large magneto-resistance phenomena in ruthenates, since 
in our phase diagram the ``metallic'' FM phase is adjacent to the
``insulating'' AFM state.
This two-phase competition is at the heart of CMR in manganites,
as has been emphasized in this chapter, and thus, CMR-like phenomenon could
also exist in ruthenates.
For more details the reader should consult Hotta and Dagotto, 2001.

\section{Phase-separation scenario}

\subsection{Phase Diagram with Classical Localized Spins}

Although the one-orbital model for manganites is clearly incomplete to 
describe these compounds since, by definition, it has only one active
orbital, nevertheless it has been shown in recent calculations
that it captures part of the interesting competition between
ferromagnetic and antiferromagnetic phases in these compounds
(see, for instance, Moreo et al., 1999a).
For this reason, and since this model is far simpler than the more 
realistic two-orbital model, it is useful to study it in detail.

\begin{figure}[t]
\centerline{\epsfxsize=8.truecm \epsfbox{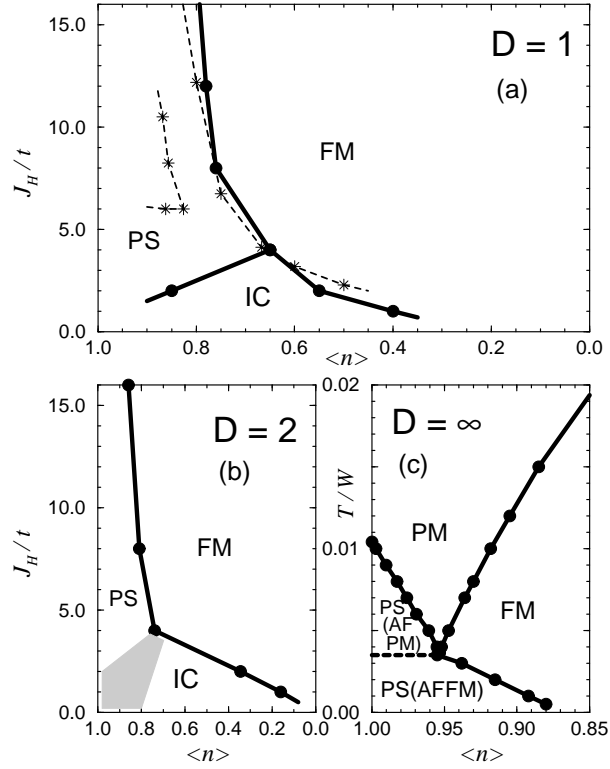}}
\caption{
Phase diagram of the one-orbital model with classical spins (and
without $J_{\rm AF}$ coupling).
(a) are results obtained with Monte Carlo methods 
at low temperature in 1D (Yunoki et al., 1998a; Dagotto
et al., 1998). FM, PS, and IC, denote ferromagnetic, phase-separated,
and spin incommensurate phases, respectively. Although not shown
explicitly, the \densi=1.0 axis is antiferromagnetic. The dashed lines
correspond to results obtained using quantum localized spins. For more
details see Yunoki et al. (1998a) and Dagotto et al. (1998).
(b) Similar to (a) but in 2D. The grey region denotes the possible 
location of the PS-IC transition at low Hund coupling, which is
difficult to determine.
Details can be found in Yunoki et al. (1998a).
(c) Results obtained in the infinite dimension limit and at large Hund 
coupling varying the temperature (here in units of the half-width
$W$ of the density of states). Two regions with PS were identified, as
well as a paramagnetic PM regime.
For details see Yunoki et al. (1998a).}
\label{fig11}
\end{figure}

A fairly detailed analysis of the phase diagram of the one-orbital
model has been recently presented, mainly using computational 
techniques.
Typical results obtained in Yunoki et al. (1998a)
are shown in Fig.~\ref{fig11}(a)-(c) for $D$=1, 2, and $\infty$
($D$ is spatial dimension), the first two obtained with Monte Carlo
techniques at low temperature, and the third with the dynamical
mean-field approximation in the large $J_{\rm H}$ limit varying
temperature.
There are several important features in the results which are common
in all dimensions.
At $e_{\rm g}$-density \densi=1.0, the system is antiferromagnetic
(although this is not clearly shown in Fig.~\ref{fig11}). 
The reason is that at large Hund coupling, double occupancy in the
ground state is negligible at $e_{\rm g}$-density \densi=1.0 
or lower, and at these densities
it is energetically better to have 
nearest-neighbor spins antiparallel, gaining an energy of order 
$t^2/J_{\rm H}$, rather than to align them, since in such a case the system
is basically frozen due to the Pauli principle.
On the other hand, at finite hole density, antiferromagnetism is
replaced by the tendency of holes to polarize the spin background to
improve their kinetic energy.
Then, a very prominent ferromagnetic phase develops in the model as 
shown in Fig.~\ref{fig11}.
This FM tendency appears in all dimensions of  interest, and it manifests
itself in the Monte Carlo simulations
through the rapid growth with decreasing temperature, and/or increasing
number of sites, of the zero-momentum spin-spin correlation, as shown
by Yunoki et al. (1998a). 
In real space, the results correspond to spin correlations between two
sites at a distance $d$ which do not decay to a vanishing number as $d$
grows, if there is long-range order
(see results in Dagotto et al., 1998).
In 1D, quantum fluctuations are expected to be so strong that
long-range order cannot be achieved, but in this case the spin 
correlations still can decay slowly with distance following a power law.
In practice, the tendency toward FM or AF is so strong even in 1D that 
issues of long-range order vs power-law decays are not of much
importance for studying the dominant tendencies in the model.
Nevertheless, care must be taken with these subtleties if very
accurate studies are attempted in 1D.

The most novel result emerging from the computational studies of the
one-orbital model is the way in which the FM phase is reached by hole
doping of the AF phase at \densi=1.0.
As explained before, mean-field approximations by de Gennes (1960)
suggested that this interpolation should proceed through a so-called
``canted'' state in which the spin structure remains antiferromagnetic 
in two directions but develops a uniform moment along the third
direction.
For many years this canted state was assumed to be correct, and many
experiments were analyzed based on such state.
However, the computational studies showed that instead of a canted
state, an electronic  ``phase separated'' (PS) regime interpolates
between the FM and AF phases. This PS region is very prominent in the
phase diagram of Fig.~\ref{fig11}(a)-(c) in all dimensions.

\begin{figure}[t]
\centerline{\epsfxsize=8.truecm \epsfbox{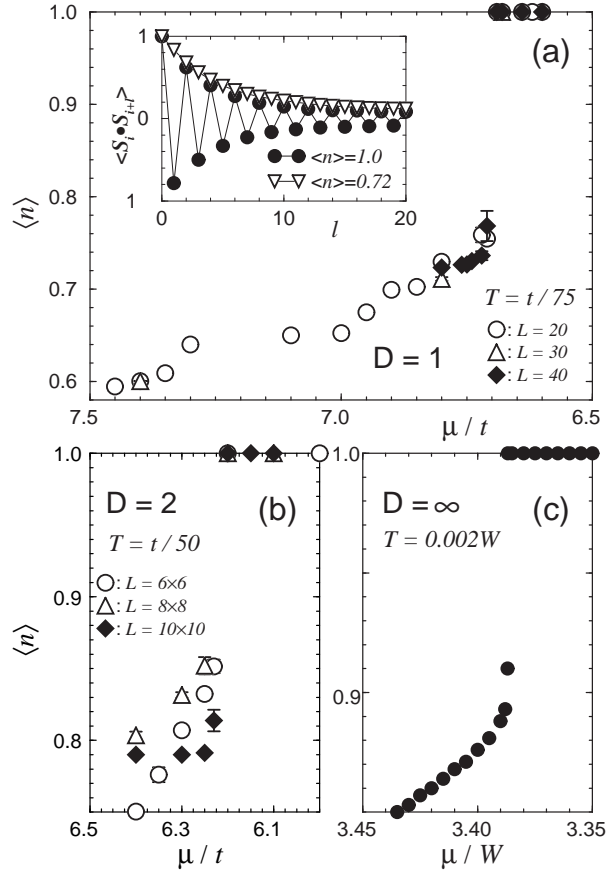}}
\caption{
Density of $e_{\rm g}$ electrons vs chemical potential $\mu$.
The coupling is $J_{\rm H}$=8$t$ in (a) and (b) and 4$W$ in (c) ($W$
is the half-width of the density of states). Temperatures and lattice
sizes are indicated.
(a) Results in 1D with PBC. The inset contains the spin correlations
at the electronic densities 1.00 and 0.72, that approximately limit
the density discontinuity. (b) Same as (a) but in 2D. (c) Same as (a)
but in D=$\infty$. Results reproduced from Yunoki et al. (1998a).}
\label{fig12}
\end{figure}

As an example of how PS is obtained from the computational work,
consider Fig.~\ref{fig12}.
In the Monte Carlo simulations carried out in this context,
performed in the grand-canonical ensemble, the density of mobile
$e_{\rm g}$-electrons \densi is an output of the calculation,
the input being the chemical potential $\mu$.
In Fig.~\ref{fig12}(a), the density \densi vs. $\mu$ is shown
for one dimensional 
clusters of different sizes at low temperature and large Hund
coupling, in part (b) results in two dimensions are presented, and in 
part (c) the limit $D$=$\infty$ is considered.
In all cases, a clear $discontinuity$ in the density appears at a
particular value of $\mu$, as in a first-order phase transition.
This means that there is a finite range of densities which are simply
unreachable, i.e., that they cannot be stabilized regardless of how
carefully $\mu$ is tuned.
In the inset of Fig.~\ref{fig12}(a), the spin correlations are shown for the
two densities at the extremes of the discontinuity, and they
correspond to FM and AF states.

Strictly speaking, the presence of PS means that the model has a range
of densities which cannot be accessed, and thus, those densities are
simply $unstable$. 
This is clarified better using now the canonical ensemble, where the
number of particles is fixed as an input and $\mu$ is an output.
In this context, suppose that one attempts to stabilize a density
such as \densi=0.95 (unstable in Fig.~\ref{fig12}),
by locating, say, 95 electrons into a 10$\times$10 lattice.
The ground-state of such a system will not develop a uniform
density, but instead two regions separated in space will be formed: a
large one with approximately 67 sites and 67 electrons (density 1.0)
and a smaller one with 33 sites and 28 electrons (density
$\sim$0.85). The last density is the lower value in the discontinuity
of Fig.~\ref{fig12}(b) in 2D, i.e., the first stable density after \densi=1.0
when holes are introduced.
Then, whether using canonical or grand-canonical approximations, a
range of densities remains unstable.

The actual spatial separation into two macroscopic regions (FM and AF
in this case) leads to an energy problem. 
In the simulations and other mean-field approximations that produce
PS, the ``tail'' of the Coulomb interaction was not explicitly
included.
In other words, the electric charge was not properly accounted for.
Once  this long-range Coulomb interaction is introduced into the
problem, the fact that the FM and AF states involved in PS have
different densities leads to a huge energy penalization 
even considering a large dielectric constant due to polarization
(charge certainly cannot be accumulated in a macroscopic 
portion of a sample).
For this reason, it is more reasonable to expect that the PS domains
will break into smaller pieces.
The shape of these pieces remains to be investigated in detail since
the calculations are difficult with long-range interactions (for
results in 1D see below), but droplets or stripes appear as a serious
possibility. 
This state would now be $stable$, since it would satisfy in part the 
tendency toward phase separation and also it will avoid a macroscopic 
charge accumulation. Although detailed calculations are not available,
the common folklore is that the typical size of the clusters in the
mixed-phase state arising from the competition PS vs. $1/r$ Coulomb
will be in the $nanometer$ scale, i.e., just a few lattice spacings 
since the Mn-Mn distance is about 4$\rm \AA$.
This is the electronic ``Phase Separated'' state that one usually has
in mind as interpolating between FM and AF.
Small clusters of FM are expected to be created in the AF background,
and as the hole density grows, these clusters will increase in number
and eventually overcome the AF clusters.

The unstable character of the low hole-density region of the phase
diagram corresponding to the one-orbital model for manganites has also
been analyzed by other authors using mostly analytic approximate 
techniques. The reader can find an extensive list of references in
the review Dagotto, Hotta, and Moreo (2001) or in Dagotto (2002).

\subsection{Electronic Phase Separation with Two Orbitals}

\begin{figure}[t]
\centerline{\epsfxsize=9.truecm \epsfbox{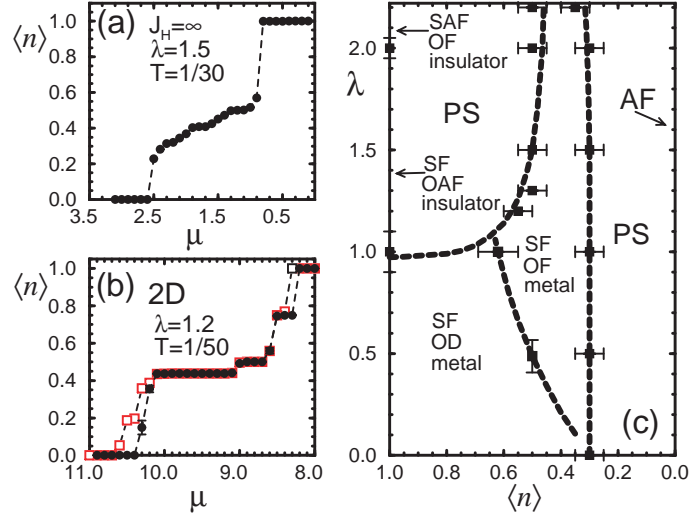}}
\caption{
(a) \densi vs $\mu$ at the couplings and temperature indicated on a
L=22 site chain. The discontinuities characteristic of phase
separation are clearly shown. (b) Same as (a) but in 2D at the
parameters indicated.
The two sets of points are obtained by increasing and decreasing
$\mu$, forming a hysteresis loop. (c) Phase diagram of the two
orbitals model in 1D, $J_{\rm H}$=8, $J'$=0.05, and using the hopping
set $t_{\rm aa}$=$t_{\rm bb}$=2$t_{\rm ab}$=2$t_{\rm ba}$. 
The notation has been explained in the text. For more details see
Yunoki et al. (1998b), from where this figure was reproduced.}
\label{fig13}
\end{figure}

Now let us analyze the phase diagram at densities away from \densi=1.0.
In the case of the one-orbital model, phase separation
was very prominent in this regime. 
In the case of the two orbitals model, discontinuities in
\densi vs. $\mu$ appear similarly as for one orbital,
signalling the presence of phase separation tendencies,
as shown in Fig.~\ref{fig13}(a) and (b).
Measurements of spin and orbital correlations, as well as the Drude
weight to distinguish between metallic and insulating behavior, have
suggested the phase diagram in one dimension reproduced
in Fig.~\ref{fig13}(c).
There are several phases in competition.
At \densi=1.0 the results were already described in the previous
subsection. Away from the \densi=1.0 phases, only the spin-FM
orbital-disordered survives at finite hole density, as expected due to
the mapping at small $\lambda$ into the one-orbital model with half
the density.
The other phases at $\lambda$$\geq$1.0 are not stable, but 
electronic phase
separation takes place.
The \densi$<$1.0 extreme of the PS discontinuity is given by a
spin-FM orbital-FM metallic state, which is a 1D precursor of the
metallic orbitally-ordered A-type state identified in some compounds 
precisely at densities close to 0.5.
Then, the two states that compete in the \densi$\sim$1.0 PS regime
differ in their orbital arrangement, but not in the spin sector.
This is PS triggered by the $orbital$ degrees of freedom, which is a
novel concept.
On the other hand, the PS observed at low density is very similar to 
that observed in the one-orbital model involving spin FM and AF states 
in competition.
Finally, at \densi$\sim$0.5 and large $\lambda$, charge ordering
takes place, but this phase will be discussed in more detail
later. Overall, it is quite reassuring to observe that the stable
phases in Fig.~\ref{fig13}(c) all have an analog in experiments.
This gives support to the models used and to the computational and
mean-field techniques employed.

In addition, since all stable regions are realistic, it is natural to
assume that the rest of the phase diagram, namely the PS regions, must 
also have an analog in experiments in the form of mixed-phase
tendencies and nanometer-size cluster formation, as discussed in the
case of the one-orbital model.
PS is very prominent in all the models studied, as long as proper
many-body techniques are employed.

\begin{figure}[t]
\centerline{\epsfxsize=8.truecm \epsfbox{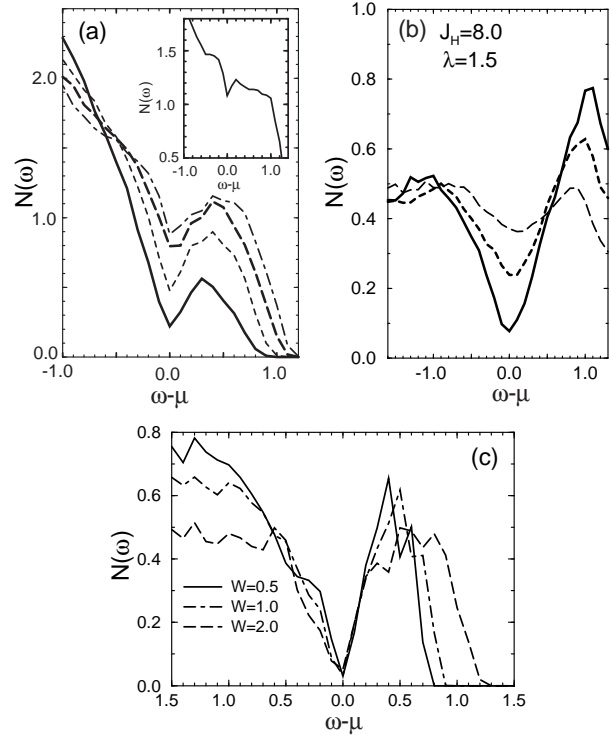}}
\caption{
(a) DOS of the one-orbital model on a 10$\times$10 cluster
at $J_{\rm H}$=$\infty$ and temperature $T$=1/30 (hopping $t$=1).
The four lines from the top correspond to densities 0.90, 0.92, 0.94,
and 0.97. The inset has results at \densi=0.86, a marginally stable
density at $T$=0. 
(b) DOS of the two-orbital model on a 20-site chain, working 
at \densi=0.7, $J_{\rm H}$=8, and $\lambda$=1.5. Starting from the 
top at $\omega$-$\mu$=0, the three lines represent temperatures
1/5, 1/10, and 1/20, respectively.
Here the hopping along $x$ between orbitals a is the unit of energy.
Both, (a) and (b) are taken from Moreo, Yunoki, and Dagotto (1999b).
(c) DOS using a 20-site chain of the one-orbital model
at $T$=1/75, $J_{\rm H}$=8, \densi=0.87, and at a chemical potential
such that the system is phase-separated in the absence of
disorder. $W$ regulates the strength of the disorder, 
as explained in Moreo et al. (2000) from where this figure was taken.}
\label{fig14}
\end{figure}

\subsection{Pseudogap in Mixed-Phase States}

Recent theoretical investigations suggest that the density of states
(DOS) in mixed-phase regimes of manganites 
may have ``pseudogap'' characteristics, 
namely a prominent depletion of weight at the chemical potential.
This feature is similar to that extensively discussed in copper
oxides. The calculations in the Mn-oxide context have been carried out
using both the one- and two-orbital models, with and without disorder 
(see Moreo, Yunoki and Dagotto, 1999b; Moreo et al., 2000).
Typical results are shown in Fig.~\ref{fig14}.
Part (a) contains the DOS of the one-orbital model on a 2D cluster
varying the electronic density slightly below \densi=1.0,
as indicated in the caption.
At zero temperature, this density regime is unstable due to phase
separation, but at the temperature of the simulation those densities
still correspond to stable states, but with a dynamical mixture of AF
and FM features (as observed, for instance, in Monte Carlo snapshots
of the spin configurations).
A clear minimum in the DOS at the chemical potential can be observed.
Very similar results appear also in 1D simulations (Moreo, Yunoki and
Dagotto, 1999b). Part (b) contains results for two-orbitals and a
large electron-phonon coupling, this time at a fixed density and changing
temperature. Clearly a pseudogap develops in the system as a precursor
of the phase separation that is reached as the temperature is further
reduced. Similar results have been obtained in other parts of
parameter space, as long as the system is near unstable
phase-separated regimes.
Pseudogaps in the DOS appear also through the influence of disorder on
first-order transitions (see Fig.~\ref{fig14}(c)), issue to be discussed
in the next section".

A tentative detailed explanation of this phenomenon for the case without
disorder was described by Moreo, Yunoki, and Dagotto (1999b), but
intuitively it is clear that a mixture of an insulator (with a gap in
the density of states) and a metal (with a featurless density of
states), will lead to an intermediate situation as it is the pseudogap.
In this respect, the appearance of the pseudogap is natural when metals
and insulators compete.

\subsection{Phase Separation Caused by the Influence of Disorder
on First-Order Transitions}

Although it is frequently stated in the literature that a variety of
chemical substitutions in manganites lead to modifications in the
bandwidth due to changes in the ``average'' A-site cation radius 
$\langle r_{\rm A} \rangle$, this statement is only partially true.
Convincing analysis of data and experiments by Rodriguez-Martinez and
Attfield (1996) have shown that the disorder introduced by chemical
replacements in the A-sites is also crucially important in determining
the properties of manganites.
For instance, Rodriguez-Martinez and Attfield (1996) found that the
critical temperature $T_{\rm C}$ can be reduced by a large factor if
the variance $\sigma^2$ of the ionic radii about the mean
$\langle r_{\rm A} \rangle$ is modified, keeping 
$\langle r_{\rm A} \rangle$ constant.
Rodriguez-Martinez and Attfield (1996) actually observed that maximum
magnetoresistance effects are found in materials not only with a low
value of $\langle r_{\rm A} \rangle$ (small bandwidth) but also a
small value of $\sigma^2$.
A good example is $\PCMO$ since the Pr$^{3+}$ and Ca$^{2+}$ ions are
similar in size (1.30 $\rm \AA$ and 1.34 $\rm \AA$, respectively,
according to Tomioka and Tokura (1999)).

Disorder, as described in the previous paragraph, is
important for the phase separation scenario.
The recent experimental results showing the existence of micrometer
size coexisting clusters in 
$\rm (La_{5/8-y} Pr_y) Ca_{3/8} Mn O_3$ (LPCMO)
by Uehara et al. (1999), to be reviewed in detail later, highlights a
property of manganites that appears universal, namely the presence of
intrinsic inhomogeneities in the system, even in single crystals. 
This issue is discussed at length in various sections of this review.
In the theoretical framework described thus far, the scenario that is
the closest to predicting such inhomogeneous state is the one based 
on electronic phase separation. However, the analysis presented before
when considering the influence of long-range Coulomb interactions over
a phase separated state, led us to believe that only nanometer size
coexisting clusters are to be expected in this problem.
Those found in LPCMO are much larger, suggesting that there must be
another mechanism operative in manganites to account for their
formation.

A possible explanation of the results of Uehara et al. (1999) has been
recently proposed by Moreo et al. (2000), and it could be considered
as a form of ``disorder-induced'' or
``structural'' phase separation, rather than electronic.
The idea is based on the influence of disorder over the first-order
metal-insulator (or FM-AF) transition found in models where the
interactions are translationally invariant (without disorder).
When such a transition occurs, abruptly a metal changes into an
insulator, as either concentrations or couplings are suitably
changed. 
Unless metastable states are considered, there is no reason to assume 
that in the actual stable ground-state of this system coexisting
clusters will be found, namely the state is entirely FM or AF
depending on parameters. 
However, different is the situation when disorder is considered into
the problem.
The type of disorder taken into account by Moreo et al. (2000) is
based on the influence of the different ionic radius of the various
elements that compose the manganites.
Depending on the environment of A-type ions (which in LPCMO involve
La, Pr or Ca) a given Mn-O-Mn bond can be straight (180$^{\circ}$) or 
distorted with an angle less than  180$^{\circ}$. 
In the latter, the hopping across the bond under study will be less
than the ideal one. 
The random character of the distribution of A
ions, leads to a concomitant random distribution of hoppings, and also 
random exchange between the localized spins $J_{\rm AF}$ since this
quantity is also influenced by the angle of the Mn-O-Mn bond.

\begin{figure}[t]
\centerline{\epsfxsize=9.truecm \epsfbox{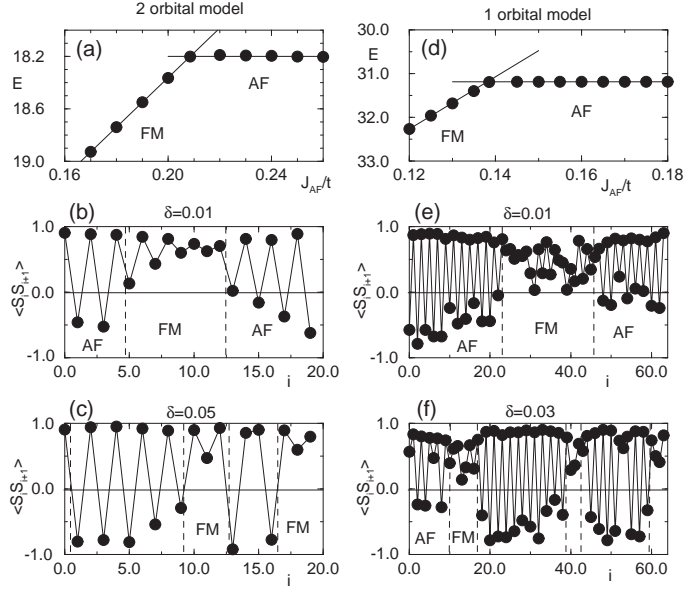}}
\caption{
Results that illustrate the generation of ``giant'' coexisting
clusters in models for manganites (taken from Moreo et al., 2000).
(a-c) are Monte Carlo results for the two-orbital model with 
$\langle n \rangle$=0.5, $T$=1/100, $J_{\rm H}$=$\infty$, 
$\lambda$=1.2, $t$=1, PBC, and using a chain with L=20 sites.
(a) is the energy per site vs $J_{\rm AF}/t$ for the non-disordered
model, with level crossing at 0.21.
(b) MC averaged nearest-neighbor $t_{\rm 2g}$-spins correlations 
vs position along the chain (denoted by i) for one set of random
hoppings $t^{\alpha}_{\rm ab}$ and $J_{\rm AF}$ couplings 
($J_{\rm AF}/t$ at every site is between 0.21-$\delta$ and
0.21+$\delta$ with $\delta$=0.01). FM and AF regions are shown. For
more details see Moreo et al. (2000). 
(c) Same as (b) but with $\delta$=0.05.
(d-f): results for the one-orbital model with
$\langle n \rangle$=0.5, $T$=1/70, $J_{\rm H}$=$\infty$, $t$=1, open
boundary conditions, and L=64 (chain). (d) is energy per site vs 
$J_{\rm AF}$ for the non-disordered model, showing the FM-AF states
level crossing at $J_{\rm AF}$$\sim$0.14. 
(e) are the MC averaged nearest-neighbor $t_{\rm 2g}$-spin
correlations vs position for one distribution of random hoppings and
$t_{\rm 2g}$ exchanges, such that $J_{\rm AF}/t$ is 
between 0.14-$\delta$ and 0.14+$\delta$ with $\delta$=0.01.
(f) Same as (e) but with $\delta$=0.03.}
\label{fig15}
\end{figure}

To account for this effect, Moreo et al. (2000) studied the one- and
two-orbital models for manganites described before, including a
small random component to both the hoppings and $J_{\rm AF}$.
This small component did not influence the FM and AF phases much
away from their transition boundary, but in the vicinity of the
first-order transition its influence is important.
In fact, numerical studies show that the transition now becomes
continuous, with FM and AF clusters in coexistence in a narrow region
around the original transition point. 

Typical results are shown in Figs.~\ref{fig15}(a)-(f), using one-dimensional
clusters as an example. In the two upper frames, the energy versus 
$J_{\rm AF}$ (or $J'$) is shown at fixed values of the other
couplings such as $J_{\rm H}$ and $\lambda$, in the absence of
disorder  and at a fixed density x=0.5. 
The abrupt change in the slope of the curves in (a) and (d) clearly
shows that the transition is indeed first-order.
This is a typical result that appears recurrently in all Monte Carlo
simulations of manganite models, namely FM and AF are so different
that the only way to change from one to the other at low temperature
is abruptly in a discontinuous transition (and spin canted 
phases have not been found in our analysis in the absence of magnetic 
fields, as possible intermediate phases between FM and AF). 
These results are drastically changed upon the application of
disorder, as shown in frames (b,c,e, and f) of Fig.~\ref{fig15},
where the
mean couplings have been fixed such that the model is located exactly
at the first-order transition of the non-disordered system.
In these frames, the nearest-neighbor spin correlations along the chain 
are shown. Clearly this correlation is positive in some portions of
the chain, while it alternates from positive to negative in
others. This alternation is compatible with an AF state, with an
elementary unit cell of spins in the configuration
 up-up-down-down, but the particular
form of the AF state is not important in the following; only its
competition with other ordered states, such as the FM one is significant.
The important point is that there are coexisting FM and AF regions.
The cluster size is regulated by the strength of the disorder, such
that the smaller the disorder, the larger the cluster size.
Results such as  those in Fig.~\ref{fig15} have appeared in all simulations
carried out in this context, and in dimensions larger than one 
(see Moreo et al., 2000).
The conclusions appear independent of the particular type of AF
insulating state competing with the FM state, the details of the
distribution of random numbers used, and the particular type of
disorder considered which could also be in the form of a random
on-site energy in some cases (Moreo et al., 2000). 
Note that the coexisting clusters have the $same$ density, namely
these are FM and AF phases that appear at a fixed hole concentration
in the non-disordered models, for varying couplings.
Then, the problem of a large penalization due to the accumulation of
charge is not present in this context.

What is the origin of such a large cluster coexistence with equal
density? 
There are two main opposing tendencies acting in the system.
On one hand, energetically it is not convenient to create FM-AF
interfaces and from this perspective a fully homogeneous system is
preferable.
On the other hand, locally at the level of the lattice spacing
the disorder in $t$ and $J_{\rm AF}$ alter
the couplings such that the system prefers to be either on the FM or
AF phases, since these couplings fluctuate around the 
transition value. From the perspective of the disorder, 
the clusters should be as
small as possible such that the local different tendencies can be
properly accounted for. From this competition emerges the large
clusters of Fig.~\ref{fig15}, namely by creating large clusters,
the number of
interfaces is kept small while the local tendencies toward one phase
or the other are partially satisfied. ``Large'' here means substantially
larger in size than the lattice spacing.
A region where accidentally the distribution of random couplings
favors the FM or AF state on average, will nucleate such a phase in
the form of a bubble.

Very recent results in this context have been presented by Burgy 
{\it et al.} (2001) where the competition of two phases was analyzed in
general terms, and the resistivity was calculated with a resistor
network approximation. A clearly large MR effect was observed, and
extensions to cuprates were proposed.

Summarizing, phase separation can be 
driven by energies other than purely electronic.
In fact it can also be triggered by the influence of
disorder on first-order transitions. In this case the competing
clusters have the same density and for this reason can be very
large. Micrometer size clusters, such as those found in the RFIM,  are
possible in this context, and have been observed 
in experiments. This result  is very
general, and should apply to a variety of compounds where two very
different ordered states are in competition at low temperatures. 

\subsection{Resistivity of Manganites in the Mixed-Phase Regime}

One of the main lessons learned from the previous analysis of models 
for manganites is that intrinsic inhomogeneities are very important 
in this context.
It is likely that the real Mn-oxides in the CMR regime are in such 
a mixed-phase state, a conclusion that appears inevitable based on 
the huge recent experimental literature, 
reporting phase separation tendencies in some form or
another in these compounds.
However, note that until recently estimations of the d.c. resistivity
$\rho_{\rm dc}$ in such a mixed-phase regime were not available. 
This was unfortunate since the interesting form of the $\rho_{\rm dc}$
vs. temperature curves, parametric with magnetic fields, is one of the
main motivations for the current huge effort in the manganite context.
However, the lack of $reliable$ estimations of $\rho_{\rm dc}$ is not 
accidental: it is notoriously difficult to calculate transport
properties in general, and even more complicated in regions of
parameter space that are expected to be microscopically inhomogeneous.
Although there have been some attempts in the literature to calculate
$\rho_{\rm dc}$, typically a variety of approximations that are not
under control have been employed. In fact, the micrometer size of some
of the coexisting clusters found in experiments strongly suggest that
a fully microscopic approach to the problem will likely fail since,
e.g., in a computational analysis it would be very difficult to study
sufficiently large clusters to account for such large scale
structures. It is clear that a more phenomenological approach is needed
in this context.

\begin{figure}[t]
\centerline{\epsfxsize=9.truecm \epsfbox{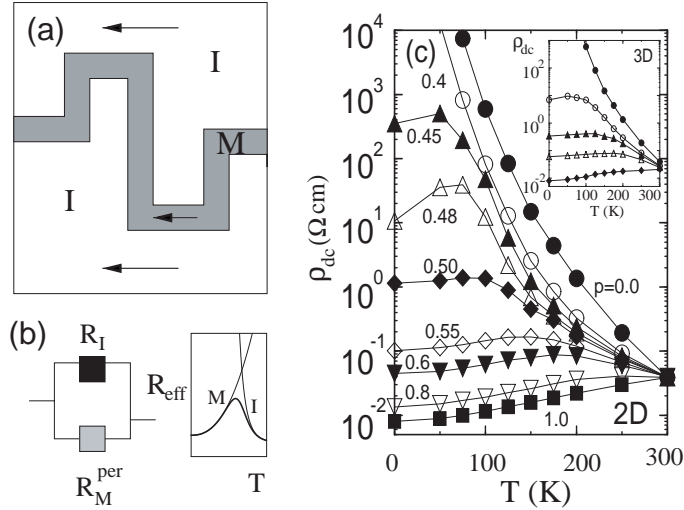}}
\caption{
(a) Mixed-phase state near percolation. Arrows indicate conduction
through insulating or metallic regions depending on T. 
(b) Two-resistances model for Mn-oxides. 
Effective resistance $\rm R_{eff}$ vs. T (schematic) 
arising from the parallel connection
of metallic (percolative) $\rm R^{per}_M$ and insulating 
$\rm R_I$ resistances. 
(c) Net resistivity $\rho_{\rm dc}$ of a 100$\times$100 random resistor
network cluster vs temperature, at the indicated metallic fractions
$p$ (result taken from Mayr et al., 2000). Inset: Results for a 20$^3$
cluster with (from the top) $p$=0.0, 0.25, 0.3, 0.4 and 0.5. 
In both cases, averages over 40 resistance configurations were
made. The $p$=1 and 0 limits are from the experiments  
corresponding to LPCMO (see Uehara et al., 1999). 
Results on 200$\times$200 clusters (not shown) indicate that size
effects are negligible. }
\label{fig16}
\end{figure}

For all these reasons, recently a two effective resistance picture of
the physics of manganites was proposed by Mayr et al. (2000).
A sketch of this idea is in Fig.~\ref{fig16}.
Mayr et al. (2001) carried out a study of $\rho_{\rm dc}$ using
a $random$ $resistor$ $network$ model 
(see Kirkpatrick, 1973), and other approximations.
This model was defined on square and cubic lattices, but with a
lattice spacing much larger than the $\rm 4\AA$ distance between
nearest-neighbor Mn ions. 
Actually, the new lattice spacing is a fraction of micrometer, since
the random network tries to mimic the complicated fractalic-like
structure found experimentally.
At each link in this sort of effective lattice, randomly either a
metallic or insulating resistance was located in such a way that the
total fraction of metallic component was $p$, a number between 0 and 1.

The actual values of these resistances as a function of temperature
were taken from experiments. Mayr et al. (2001) used the
$\rho_{\rm dc}(T)$ plots obtained by Uehara et al. (1999)
corresponding to $\rm (La_{5/8-y} Pr_y) Ca_{3/8} Mn O_3$ (LPCMO), one
of the compounds that presents the coexistence of giant FM and CO
clusters at intermediate values of the Pr concentration. More
specifically, using for the insulating resistances the results of
LPCMO at y=0.42 (after the system becomes a CO state with 
increasing Pr doping) and for the metallic ones the results at y=0.0
(which correspond to a metallic state, at least below 
its Curie temperature), the results of a numerical study 
on a 100$\times$100 cluster
are shown in Fig.~\ref{fig16}
(the Kirchoff equations were solved by a
simple iterative procedure).
It is interesting to observe that, even using such a simple
phenomenological model, the results are already in reasonable
agreement with the experiments, namely, (i) at large temperature
insulating behavior is observed even for $p$ as large as 0.65 (note
that the classical percolation is expected to occur near $p=0.5$; see 
Kirkpatrick (1973)); (ii) at small temperature a (``bad'') metallic
behavior appears; and (iii) a broad peak exists in between.
Results in both 2D and 3D lead to similar conclusions. It is clear
that the experimental results for manganites can be at least partially
accounted for within the mixed-phase scenario.

The results of Fig.~\ref{fig16} suggest a simple qualitative picture
to visualize why the resistivity in Mn-oxides has the peculiar shape it
has. The relevant state in this context should be imagined as
percolated.  
Metallic filaments from one side of the sample to the other exist 
in the system. At low temperature, conduction is through those
filaments. 
Necessarily, $\rho_{\rm dc}$ at $T$=0 must be large, in such a
percolative regime.
As temperature increases, the $\rho_{\rm dc}$ of the filaments grows as
in any metal. However, in the other limit of large or room temperature,
the resistance of the percolated metallic filament is expected to be
much larger than
that corresponding to one of the insulator paths. Actually, near room
temperature in many experimental graphs, it can be observed that
$\rho_{\rm dc}$ in the metallic and insulating regimes are quite
similar in value, even comparing
results away from the percolative region.
Then, at room temperature it is more likely that conduction will occur
through the insulating portions of the sample, rather than through the
metallic filaments. Thus, near room temperature insulating
behavior is expected. In between low and high temperature, it is
natural that $\rho_{\rm dc}$ will present a peak. Then, a simple 
``two resistances in parallel'' description appears appropriate. 
The insulating resistance behaves like any insulator,
while the metallic one starts at $T$=0 at a high value and then it
behaves like any metal. The resulting effective resistance 
properly reproduces the experiments at least qualitatively.

The success of the phenomenological approach described above leads to
an interesting prediction. In the random resistor network, it is clear
that above the peak in the resistivity, the mixed-phase character of
the system remains, even with a temperature dependent metallic
fraction $p$.
Then, it is conceivable to imagine that above the Curie temperature in
real manganites, a substantial fraction of the system should remain in
a metallic FM state (likely not percolated, but forming disconnected
clusters). 
A large variety of experiments indeed
suggest that having FM clusters above $T_{\rm C}$ is possible. As a
consequence, this has led us to conjecture that there must exist a
temperature $T^*$ at which those clusters start forming. This defines
a new temperature scale in the problem, somewhat similar to the famous
pseudogap $T^*$ scale of the high temperature superconducting
compounds. 
In fact, in mixed phase FM-AF states it is known that a pseudogap
appears in the density of states (Moreo et al., 1999b;
Moreo et al., 2000), thus increasing the analogy between these two
materials. In our opinion, the experimental verification that indeed
such a new scale $T^*$ exists in manganites is important to
our understanding of these compounds. In fact, 
recent results by Kim, Uehara and
Cheong (2000) for $\LCMO$ at various densities have been interpreted
as caused by small FM segments of the CE-type CO state,
appearing at hole densities smaller than x=1/2 and at high temperature.
This result is in qualitative agreement with the theoretical
analysis presented here.

The study of effective resistivities and conductances has also been 
carried out in the presence of magnetic fields (Mayr et al., 2001),
although still mainly within a phenomenological approach. From the
previous results, it is clear that in the percolative
regime ``small'' changes in the system may lead to large changes in
the resistivity. 
It is conceivable that small magnetic fields could induce
such small changes in $p$, leading to substantial modifications in the
resistivity.
Experiments by Parisi et al. (2001) indeed show a rapid change of
the fraction of the FM phase in $\rm La_{0.5} Ca_{0.5} Mn O_3$
upon the application of magnetic fields.
In addition, studies of the one-orbital model carried out
in one dimension (Mayr et al., 2001) also showed that other factors
may influence the large $\rho_{\rm dc}$ changes upon the
application of external fields.

\section{Concluding Remarks}

In this review, a small fraction of the main results gathered in recent years
in the context of theoretical studies of models for manganites have been
discussed.
For better information the reader should consult a more extensive review
by the authors and A. Moreo (Dagotto, Hotta, and Moreo, 2001)
or the recent book on manganites by Dagotto (Dagotto 2002).
The main experiments that have helped clarify the physics of these
interesting compounds are reviewed elsewhere in this book. 
Several aspects of the problem are by now widely accepted,
while others still need further work to be confirmed. Intrinsic
inhomogeneities exist in models and experiments and seem to play
a key role in these compounds. 

Among the issues related with inhomogeneities that after a
considerable effort appear well-established are the following:

\noindent {\bf (1)} 
Work carried out by several groups using a variety of techniques
have shown that electronic phase separation is a dominant feature of 
models for manganites, particularly in the limits of small and large 
hole doping. This type of phase separation leads to nanometer size
coexisting clusters once the long-range Coulombic repulsion is
incorporated into the models.

\noindent {\bf (2)} 
Working at constant density, the transitions between metallic
(typically FM) and insulating (typically CO/AF) states are of 
{\bf first} order at zero temperature. No counter-example has been
found to this statement thus far.

\noindent {\bf (3)}
A second form of phase separation has been recently discussed.
It is produced by the influence of disorder on the first-order 
metal-insulator transitions described in the previous item.
If couplings are fixed such
that one is exactly at the first-order 
transition in the absence of disorder, 
the system is ``confused'' and does not know whether to
be metallic or insulating (at zero disorder). On the other hand, if
the couplings are the same, but the strength of disorder is large in such
a way that it becomes dominating, then tiny clusters of the two competing
phases are formed with the lattice spacing as the typical length scale.
For nonzero but weak disorder, an intermediate situation develops where
fluctuations in the disorder pin either one phase or the other in large
regions of space. 

This form of phase separation is even more promising than the
electronic one for explaining the physics of manganites for a variety of
reasons: (i) it involves phases with the same density, thus there are
no constraints on the size of the coexisting clusters which can be as
large as a micrometer in scale, as found in experiments. 
(ii) The clusters are randomly distributed and have fractalic shapes,
leading naturally to {\bf percolative} transitions 
from one competing phase to
the other, as couplings or densities are
varied. This is in agreement with many experiments that have reported
percolative features in manganites.
(iii) The resistivity obtained in this context is similar to that
found in experiments. Near the critical amount
of metallic fraction for percolation, at room temperature the charge
conduction can occur through the insulating regions since their
resistivity at that temperature is very similar to that of the
metallic state. Thus, the system behaves as an insulator. However, at
low temperatures, the insulator regions have a huge resistivity and,
thus, conduction is through the percolative metallic filaments which
have a large intrinsic resistivity. The system behaves as a bad metal,
and $\rho_{\rm dc}(T=0)$ can be very large.
(iv) Finally, it is expected that in a percolative regime there must
be a high sensitivity to magnetic fields and other naively ``small''  
perturbations, since tiny changes in the metallic fraction can induce
large conductivity modifications. This provides the best explanation
of the CMR effect of which these authors are aware.

\noindent {\bf (4)} 
The experimental evidence for inhomogeneities in manganites is by now
simply overwhelming. Dozens of groups, using a variety of techniques, 
have converged to such a conclusion. 
It is clear that homogeneous descriptions of manganites in the region
of interest for the CMR effect are incorrect.
These inhomogeneities appear even above the Curie temperature. In
fact, the present authors believe that a new scale of temperature
$T^*$ should be introduced (see Burgy et al., 2001).
There must be a temperature window where coexisting clusters exist
above the temperatures where truly long-range order develops. Part of
the clusters can be metallic, and their percolation may induce
long-range order as temperature decreases.  The region below $T^*$ can
be as interesting as that observed in high temperature
superconductors, at temperatures higher than the critical values. It
is likely that it contains pseudogap characteristics, due to its low
conductivity in low bandwidth manganites. The search for a
characterization of $T^*$ should be pursued actively in experiments.

\noindent {\bf (5)} 
The famous CE-state of half-doped manganites has been shown to be
stable in mean-field and computational studies of models for
manganites. Although such a state was postulated a long time ago, it is
only recently that it has emerged from unbiased studies. The simplest
view to understand the  CE-state is based on a ``band insulating''
picture: it has been shown that in a zigzag FM chain a gap opens at
x=0.5, reducing the energy compared with straight chains. Thus,
elegant geometrical arguments are by now available to understand the
origin of the naively quite complicated CE-state of manganites. Its
stabilization can be rationalized based simply on models of
non-interacting spinless fermions in 1D geometries.
In addition, theoretical  studies have allowed one to analyze the
properties of the states competing with the CE at x=0.5. In order to
arrive at the CE-state, the use of a strong long-range Coulomb
interaction to induce the staggered charge pattern is not correct,
since by this procedure the experimentally observed charge-stacking
along the $z$-axis could not be reproduced, and in addition the
metallic regimes at x=0.5 found in some manganites would not be
stable. Manganese oxides are in the subtle regime where many different
tendencies are in competition.

\noindent {\bf (6)} 
Contrary to what was naively believed until recently, studies with
strong electron Jahn-Teller phonon coupling or with strong on-site
Coulomb interactions lead to quite similar phase diagrams. The reason
is that both  interactions disfavor double occupancy of a given
orbital. Thus, if the goal is to understand the CMR effect, the
important issue is not whether the material is Jahn-Teller or Coulomb
dominated, but how the metallic and insulating phases, of whatever
origin, compete. Calculations with Jahn-Teller phonons are the
simplest in practice, and they have led to phase diagrams that contain
most of the known phases found experimentally for manganites, such as
the A-type AF insulating state at x=0, the A-type AF metallic state at
x=0.5, the CE-state at x=0.5, etc.
Such an agreement theory-experiment is quite remarkable and encouraging.

\noindent {\bf (7)} 
Also contrary to naive expectations, the smallest parameter in
realistic Hamiltonians for Mn-oxides, namely ``$J_{\rm AF}$'' between
localized $t_{\rm 2g}$ spins, plays an important role in stabilizing
the experimentally observed phases of manganites, including the
CE-state. Modifications of this coupling due to disorder are as
important as those in the hopping amplitudes for $e_{\rm g}$ electron
movement.

As a conclusion, it is clear that the present prevailing paradigm for
manganites relies on a phase-separated view of the dominant state, as
suggested by dozens of experiments and also by theoretical
calculations once powerful many-body techniques are used to study
realistic models.

Although considerable progress has been achieved in recent years in
the analysis of manganites, both in theoretical and experimental
aspects, there are still a large number of issues that require further
work. Here a partial list of {\bf open questions} is included.

\noindent {\bf (a)} 
The phase separation scenario needs further experimental
confirmation. Are there counterexamples of compounds where CMR
occurs but the system appears homogeneous? 

\noindent {\bf (b)} 
On the theory front, a phase-separated percolative state is an
important challenge to our computational abilities. Is it possible to
produce simple formulas with a small number of parameters that
experimentalists can use in order to fit their, e.g., transport data?

\noindent {\bf (c)}
The generation of a ``Quantum Critical Point'' (QCP) is likely in the
context of competing phases, and preliminary results support this view 
(Burgy et al., 2001). Can this be observed experimentally?

\noindent {\bf (d)} 
There is not much reliable theoretical work carried out in the
presence of magnetic fields addressing directly the CMR effect. 

\noindent {\bf (e)} If the prediction of a phase-separated state in
the CMR regime of manganites is experimentally fully confirmed, what
are the differences between that state and a canonical ``spin-glass''?
Both share complexity and complicated time dependences, but are they
in the same class? Stated in more exciting terms, can the
phase-separated regime of manganites be considered a ``new'' state 
of matter in any respect?

\noindent {\bf (f)} 
Considerable progress has been achieved in understanding the x=0 and
x=0.5 charge/orbital/spin order states of manganites. But little is
known about the ordered states at intermediate densities, 
both in theory and experiments. Are
there stripes in manganites at those intermediate hole densities as
recently suggested by experimental and theoretical work? 

Summarizing, the study of manganites continues challenging our
understanding of transition-metal-oxides. While considerable progress
has been achieved in recent years, much work remains to be done. In
particular, a full understanding of the famous CMR effect is still
lacking, although evidence is accumulating that it may be caused by
intrinsic tendencies toward inhomogeneities in Mn-oxides and other
compounds.  Work in this challenging area of research should continue
at its present fast pace.

\section*{Acknowledgement}

The authors would like to thank 
C. Buhler, J. Burgy, S. Capponi, A. Feiguin, N. Furukawa, K. Hallberg, J. Hu,
H. Koizumi, A. Malvezzi, M. Mayr, D. Poilblanc, J. Riera, Y. Takada,
J. A. Verges, and S. Yunoki for valuable collaborations on manganites.
E. D. was supported by NSF grant DMR-0122523.
T.H. was supported by the Grant-in-Aid for Encouragement of Young 
Scientists from the Ministry of Education, Science, Sports, 
and Culture (ESSC). 
He has been also supported by the Grant-in-Aid for Scientific Research
Priority Area from ESSC
and for Scientific Research from Japan Society for the Promotion of Science.

\section*{References}
\parskip5pt

\noindent C. P. Adams, J. W. Lynn, Y. M. Mukovskii, A. A. Arsenov,
and D. A. Shulyatev,
Phys. Rev. Lett. {\bf 85}, 2553 (2000).

\noindent P. B. Allen and V. Perebeinos,
Phys. Rev. B{\bf 60}, 10747 (1999).

\noindent J. L. Alonso, L. A. Fern\'andez, F. Guinea, V. Laliena, and 
V. Mart\'in-Mayor, 2001a,
Phys. Rev. B{\bf 63}, 64416 (2001).

\noindent J. L. Alonso, L. A. Fern\'andez, F. Guinea, V. Laliena, and 
V. Mart\'in-Mayor, 2001b,
Phys. Rev. B{\bf 63}, 054411 (2001).

\noindent J. L. Alonso, L. A. Fern\'andez, F. Guinea, V. Laliena, and 
V. Mart\'in-Mayor, 2001c,
Nucl. Phys. B{\bf 596}, 587 (2001).

\noindent P. W. Anderson and H. Hasegawa,
Phys. Rev. {\bf 100}, 675 (1955).

\noindent V. I. Anisimov, I. S. Elfimov, M. A. Korotin, and K. Terakura,
Phys. Rev. B{\bf 55}, 15494 (1997).

\noindent T. Arima, Y. Tokura, and J. B. Torrance,
Phys. Rev. B{\bf 48}, 17006 (1993).

\noindent P. Benedetti and R. Zeyher,
Phys. Rev. B{\bf 59}, 9923 (1999).

\noindent J. J. Betouras and S. Fujimoto,
Phys. Rev. B{\bf 59}, 529 (1999).

\noindent A. Bocquet, T. Mizokawa, T. Saitoh, H. Namatame, and A. Fujimori,
Phys. Rev. B{\bf 46}, 3771 (1992).

\noindent M. Braden, G. Andr\'e, S. Nakatsuji, and Y. Maeno,
Phys. Rev. B{\bf 58}, 847 (1998).

\noindent C. Buhler, S. Yunoki and A. Moreo, 
Phys. Rev. Lett. {\bf 84}, 2690 (2000).

\noindent J. Burgy, M. Mayr, V. Martin-Mayor, A. Moreo, and E. Dagotto,
Phys. Rev. Lett. {\bf 87}, 277202 (2001).

\noindent M. Capone, D. Feinberg, and M. Grilli, 
Euro. Phys. J. B{\bf 17}, 103 (2000).

\noindent C. Castellani, C. R. Natoli, and J. Ranninger,
Phys. Rev. {\bf 18}, 4945 (1978).

\noindent E. Dagotto,
Rev. Mod. Phys. {\bf 66}, 763 (1994).

\noindent E. Dagotto, S. Yunoki, A. L. Malvezzi, A. Moreo, J. Hu,
S. Capponi, D. Poilblanc, and N. Furukawa, 
Phys. Rev. B{\bf 58}, 6414 (1998).

\noindent E. Dagotto, T. Hotta, and A. Moreo,
Phys. Rep. {\bf 344}, 1 (2001). 

\noindent E. Dagotto,
{\it Nanoscale Phase Separation and Colossal Magnetoresistance},
Springer-Verlag, Berlin, 2002.

\noindent P. Dai, J. A. Fernandez-Baca, N. Wakabayashi, E. W. Plummer, 
Y. Tomioka, and Y. Tokura,
Phys. Rev. Lett. {\bf 85}, 2553 (2000).

\noindent V. Emery, S. Kivelson, and H. Lin,
Phys. Rev. Lett. {\bf 64}, 475 (1990).

\noindent P. G. de Gennes,
Phys. Rev. {\bf 118}, 141 (1960).

\noindent D. S. Dessau, T. Saitoh, C.-H. Park, Z.-X. Shen,
P. Villella, N. Hamada, Y. Moritomo, and Y. Tokura, 
Phys. Rev. Lett. {\bf 81}, 192 (1998).

\noindent D. S. Dessau and Z.-X. Shen,
Chap.~5, {\it Colossal Magnetoresistance Oxides},
edited by Y. Tokura, Gordon \& Breach, New York, 1999.

\noindent L. Dworin and A. Narath,
Phys. Rev. Lett. {\bf 25}, 1287 (1970).

\noindent R. Fr\'esard and G. Kotliar, 
Phys. Rev. B{\bf 56}, 12909 (1997). 

\noindent N. Furukawa,
J. Phys. Soc. Jpn. {\bf 63}, 3214 (1994).

\noindent M. Gerloch and R. C. Slade,
{\it Ligand-Field Parameters}, Cambridge, London, 1973.

\noindent J. Goodenough,
Phys. Rev. {\bf 100}, 564 (1955).

\noindent J. S. Griffith, 
{\it The Theory of Transition-Metal Ions}, Cambridge, London, 1961.

\noindent T. Hotta, Y. Takada, and H. Koizumi,
Int. J. Mod. Phys. B{\bf 12}, 3437 (1998).

\noindent T. Hotta, S. Yunoki, M. Mayr, and E. Dagotto, 
Phys. Rev. B{\bf 60}, R15009 (1999).

\noindent T. Hotta, Y. Takada, H. Koizumi, and E. Dagotto,
Phys. Rev. Lett. {\bf 84}, 2477 (2000).

\noindent T. Hotta and E. Dagotto,
Phys. Rev. B{\bf 61}, R11879 (2000).

\noindent T. Hotta, A. Malvezzi, and E. Dagotto,
Phys. Rev. B{\bf 62}, 9432 (2000).

\noindent T. Hotta, A. Feiguin, and E. Dagotto, 
Phys. Rev. Lett. {\bf 86}, 4922 (2001).

\noindent T. Hotta and E. Dagotto,
Phys. Rev. Lett. {\bf 88}, 017201 (2002).

\noindent M. N. Iliev, M. V. Abrashev, H.-G. Lee, Y. Y. Sun,
C. Thomsen, R. L. Meng, and C. W. Chu,
Phys. Rev. B{\bf 57}, 2872 (1998).

\noindent S. Ishihara, J. Inoue, and S. Maekawa, 
Phys. Rev. B{\bf 55}, 8280 (1997).

\noindent G. Jackeli, N. B. Perkins, and N. M. Plakida,
Phys. Rev. B{\bf 62}, 372 (2000).

\noindent H. A. Jahn and E. Teller,
Proc. Roy. Soc. London A {\bf 161}, 220 (1937).

\noindent R. Kajimoto, H. Yoshizawa, H. Kawano, H. Kuwahara,
Y. Tokura, K. Ohoyama, and M. Ohashi, 
Phys. Rev. B{\bf 60}, 9506 (1999).

\noindent J. Kanamori,
J. Appl. Phys. Suppl. {\bf 31}, 14S (1960).

\noindent J. Kanamori,
Prog. Theor. Phys. {\bf 30}, 275 (1963).

\noindent K. H. Kim, M. Uehara, and S-W. Cheong,
Phys. Rev. B{\bf 62}, R11945 (2000). 

\noindent S. Kirkpatrick,
Rev. Mod. Phys. {\bf 45}, 574 (1973).

\noindent H. Koizumi, T. Hotta, Y. Takada,
1998a, Phys. Rev. Lett. {\bf 80}, 4518 (1998).

\noindent H. Koizumi, T. Hotta, Y. Takada,
1998b, Phys. Rev. Lett. {\bf 81}, 3803 (1998).

\noindent W. Koshibae, Y. Kawamura, S. Ishihara, S. Okamoto, J. Inoue,
and S. Maekawa,
J. Phys. Soc. Jpn. {\bf 66}, 957 (1997).

\noindent M. Korotin, T. Fujiwara, and V. Anisimov,
Phys, Rev. B{\bf 62}, 5696 (2000).

\noindent M. Kubota, Y. Oohara, H. Yoshizawa, H. Fujioka, K. Shimizu,
K. Hirota, Y. Moritomo, and Y. Endoh,
J. Phys. Soc. Jpn. {\bf 69}, 1986 (2000).

\noindent K. I. Kugel and D. I. Khomskii, 
Sov. Phys.-JETP {\bf 37}, 725 (1974).

\noindent J. D. Lee and B. I. Min, 
Phys. Rev. B{\bf 55}, R14713 (1997).

\noindent A. Machida, Y. Moritomo, and A. Nakamura,
Phys. Rev. B{\bf 58}, R4281 (1998).

\noindent Y. Maeno, T. M. Rice, and M. Sigrist,
Physics Today {\bf 54}, 42 (2001).

\noindent R. Maezono, S. Ishihara, and N. Nagaosa, 
Phys. Rev. B{\bf 58}, 11583 (1998).

\noindent R. Maezono and N. Nagaosa,
Phys. Rev. B{\bf 62}, 11576 (2000).

\noindent R. Mathieu, P. Svedlindh, and P. Nordblad,
Europhys. Lett. {\bf 52} 441 (2000).

\noindent M. Mayr, A. Moreo, J. Verg\'es, J. Arispe, A. Feiguin,
and E. Dagotto, 
Phys. Rev. Lett. {\bf 86}, 135 (2001).

\noindent A. Millis, B. I. Shraiman, and P. B. Littlewood,
Phys. Rev. Lett. {\bf 74}, 5144 (1995).

\noindent A. J. Millis, B. I. Shraiman, and R. Mueller,
Phys. Rev. Lett. {\bf 77}, 175 (1996)

\noindent A. J. Millis, R. Mueller, and B. I. Shraiman,
Phys. Rev. B{\bf 54}, 5405 (1996).

\noindent A. J. Millis,
Nature {\bf 392}, 147 (1998).

\noindent A. J. Millis,
Phys. Rev. Lett. {\bf 80}, 4358 (1998). 

\noindent T. Mizokawa and A. Fujimori,
Phys. Rev. B{\bf 51}, R12880 (1995). 

\noindent T. Mizokawa and A. Fujimori,
Phys. Rev. B{\bf 54}, 5368 (1996). 

\noindent T. Mizokawa and A. Fujimori,
Phys. Rev. B{\bf 56}, R493 (1997). 

\noindent T. Mizokawa, D. I. Khomskii, and G. A. Sawatzky,
Phys. Rev. B{\bf 61}, R3776 (2000). 

\noindent T. Mizokawa, L. H. Tjeng, G. A. Sawatzky, G. Ghiringhelli,
O. Tjernberg, N. B. Brookes, H. Fukazawa, S. Nakatsuji, and Y. Maeno,
Phys. Rev. Lett. {\bf 87}, 077202 (2001).

\noindent A. Moreo, S. Yunoki and E. Dagotto,
1999a, Science {\bf 283}, 2034 (1999). 

\noindent A. Moreo, S. Yunoki, and E. Dagotto,
1999b, Phys. Rev. Lett. {\bf 83}, 2773 (1999).

\noindent A. Moreo, M. Mayr, A. Feiguin, S. Yunoki and E. Dagotto,
Phys. Rev. Lett. {\bf 84}, 5568 (2000).

\noindent S. Mori, C. H. Chen, and S.-W. Cheong,
1998a, Nature {\bf 392}, 473 (1998).

\noindent S. Mori, C. H. Chen, and S.-W. Cheong,
1998b, Phys. Rev. Lett. {\bf 81}, 3972 (1998).

\noindent Y. Moritomo, Y. Tomioka, A. Asamitsu, Y. Tokura, and Y. Matsui,
Phys. Rev. B{\bf 51}, 3297 (1995).

\noindent Y. Motome and N. Furukawa, 
J. Phys. Soc. Jpn. {\bf 68}, 3853 (1999).

\noindent Y. Motome and N. Furukawa,
J. Phys. Soc. Jpn. {\bf 69}, 3785 (2000).

\noindent E. M\"uller-Hartmann and E. Dagotto, 
Phys. Rev. B{\bf 54}, R6819 (1996).

\noindent S. Nakatsuji and Y. Maeno,
Phys. Rev. Lett. {\bf 84}, 2666 (2000).

\noindent Y. Okimoto, T. Katsufuji, T. Ishikawa, A. Urushibara, 
T. Arima, and Y. Tokura,
Phys. Rev. Lett. {\bf 75}, 109 (1995).

\noindent F. Parisi, P. Levy, G. Polla, and D. Vega,
Phys. Rev. B{\bf 63}, 144419 (2001).

\noindent J.-H. Park, C. T. Chen, S.-W. Cheong, W. Bao, G. Meigs,
V. Chakarian, and Y. U. Idzerda,
Phys. Rev. Lett. {\bf 76}, 4215 (1996).

\noindent T. G. Perring, G. Aeppli, Y. Tokura, 
Phys. Rev. Lett. {\bf 80}, 4359 (1998).

\noindent W. H. Press, S. A. Teukolsky, W. T. Vitterling, and B. P. Flannery,
{\it Numerical Recipes}, Cambridge University Press, New York, 1986.

\noindent M. Quijada, J. Cerne, J. R. Simpson, H. D. Drew, K. H. Ahn,
A. J. Millis, R. Shreekala, R. Ramesh, M. Rajeswari, and T. Venkatesan,
Phys. Rev. B{\bf 58}, 16093 (1998).

\noindent P. G. Radaelli, D. E. Cox, L. Capogna, S.-W. Cheong, and
M. Marezio,
Phys. Rev. B{\bf 59}, 14440 (1999).

\noindent L. M. Rodriguez-Martinez and J. P. Attfield, 
Phys. Rev. B{\bf 54}, R15622 (1996).

\noindent T. Saitoh, A. Bocquet, T. Mizokawa, H. Namatame,
A. Fujimori, M. Abbate, Y. Takeda, and M. Takano, 
Phys. Rev. B{\bf 51}, 13942 (1995).

\noindent S. Satpathy, Z. S. Popovic, and F. R. Vukajlovic,
Phys. Rev. Lett. {\bf 76}, 960 (1996)

\noindent J. C. Slater and G. F. Koster,
Phys. Rev. {\bf 94}, 1498 (1954).

\noindent I. Solovyev, N. Hamada, and K. Terakura,
Phys. Rev. Lett. {\bf 76}, 4825 (1996).

\noindent H. Tang, M. Plihal, and D. L. Mills, 
J. Magn. Magn. Mat. {\bf 187}, 23 (1998).

\noindent Y. Tokura,
Chap.~1 of {\it Colossal Magnetoresistance Oxides},
edited by Y. Tokura, Gordon \& Breach, New York, 2000.

\noindent Y. Tomioka and Y. Tokura,
Chap.~8, {\it Colossal Magnetoresistance Oxides},
edited by Y. Tokura,  Gordon \& Breach, New York, 2000.

\noindent J. M. Tranquada, B. J. Sternlieb, J. D. Axe,
Y. Nakamura, and S. Uchida, Nature {\bf 375}, 561 (1995).

\noindent M. Uehara, S. Mori, C. H. Chen, and S.-W. Cheong,
Nature {\bf 399}, 560 (1999).

\noindent J. van den Brink, G. Khaliullin, and D. Khomskii,
Phys. Rev. Lett. {\bf 83}, 5118 (1999).

\noindent J. van den Brink, P. Horsch, F. Mack, and A. M. Ole\'s,
Phys. Rev. B{\bf 59}, 6795 (1999).

\noindent J. van den Brink and D. Khomskii,
Phys. Rev. B {\bf 63}, R140416 (2001)

\noindent L. Vasiliu-Doloc, S. Rosenkranz, R. Osborn, S. K. Sinha,
J. W. Lynn, J. Mesot, O. H. Seeck, G. Preosti, A. J. Fedro, and
J. F. Mitchell, 
Phys. Rev. Lett. {\bf 83}, 4393 (1999).

\noindent E. O. Wollan and W. C. Koehler, 
Phys. Rev. {\bf 100}, 545 (1955).

\noindent H. Yi, J. Yu, and S.-I. Lee,
Phys. Rev. B{\bf 61}, 428 (2000)

\noindent K. Yoshida, {\it Theory of Magnetism},
Springer-Verlag, Berlin, 1996.

\noindent S. Yunoki, J. Hu, A. Malvezzi, A. Moreo, N. Furukawa, and
E. Dagotto, 1998a,
Phys. Rev. Lett. {\bf 80}, 845.

\noindent S. Yunoki, A. Moreo, and E. Dagotto, 1998b,
Phys. Rev. Lett. {\bf 81}, 5612.

\noindent S. Yunoki, T. Hotta, and E. Dagotto, 
Phys. Rev. Lett. {\bf 84}, 3714 (2000).

\noindent C. Zener,
1951, Phys. Rev. {\bf 82}, 403 (1951).

\end{document}